\documentclass[
superscriptaddress,
 amsmath,amssymb,
 aps,
]{revtex4-2}

\usepackage{amssymb,amsfonts,amsmath}
\usepackage{graphicx}
\expandafter\let\csname equation*\endcsname\relax
\expandafter\let\csname endequation*\endcsname\relax

\newcommand{\comment}[1]{}

\begin{document}

\title{Spatio-temporal sampling of near-petahertz vortex fields}

\author{Johannes Bl{\"o}chl}
\affiliation{Department of Physics, Ludwig-Maximilians-Universit\"at Munich, D-85748 Garching, Germany}
\affiliation{Max Planck Institute of Quantum Optics, D-85748 Garching, Germany}
\author{Johannes Sch{\"o}tz}
\affiliation{Department of Physics, Ludwig-Maximilians-Universit\"at Munich, D-85748 Garching, Germany}
\affiliation{Max Planck Institute of Quantum Optics, D-85748 Garching, Germany}
\author{Ancyline Maliakkal} %need to check current affiliation
\affiliation{Department of Physics, Ludwig-Maximilians-Universit\"at Munich, D-85748 Garching, Germany}
\affiliation{Max Planck Institute of Quantum Optics, D-85748 Garching, Germany}
\author{Natālija Šreibere} %need to check current affiliations
\affiliation{Department of Physics, Ludwig-Maximilians-Universit\"at Munich, D-85748 Garching, Germany}
\affiliation{Max Planck Institute of Quantum Optics, D-85748 Garching, Germany}
\author{Zilong Wang}
\affiliation{Department of Physics, Ludwig-Maximilians-Universit\"at Munich, D-85748 Garching, Germany}
\affiliation{Max Planck Institute of Quantum Optics, D-85748 Garching, Germany}
\author{Philipp Rosenberger}
\affiliation{Department of Physics, Ludwig-Maximilians-Universit\"at Munich, D-85748 Garching, Germany}
\affiliation{Max Planck Institute of Quantum Optics, D-85748 Garching, Germany}
\author{Peter Hommelhoff}
\affiliation{Laser Physics, Department of Physics, Friedrich-Alexander-Universit\"at Erlangen-N\"urnberg, D-91058 Erlangen, Germany}
\author{Andre Staudte}
\affiliation{Joint Attosecond Science Laboratory, National Research Council of Canada and University of Ottawa, Ottawa, Ontario K1A0R6, Canada}
\author{Paul B. Corkum}
\affiliation{Joint Attosecond Science Laboratory, National Research Council of Canada and University of Ottawa, Ottawa, Ontario K1A0R6, Canada}
\author{Boris Bergues}
\affiliation{Department of Physics, Ludwig-Maximilians-Universit\"at Munich, D-85748 Garching, Germany}
\affiliation{Max Planck Institute of Quantum Optics, D-85748 Garching, Germany}
\author{Matthias F. Kling}
\email{kling@stanford.edu}
\affiliation{Department of Physics, Ludwig-Maximilians-Universit\"at Munich, D-85748 Garching, Germany}
\affiliation{Max Planck Institute of Quantum Optics, D-85748 Garching, Germany}
\affiliation{SLAC National Accelerator Laboratory, Menlo Park, CA 94025, USA}
\affiliation{Applied Physics Department, Stanford University, Stanford, CA 94305, USA}

\begin{abstract} 
Measuring the field of visible light with high spatial resolution has been challenging, as many established methods only detect a focus-averaged signal.
Here, we introduce a near-field method for optical field sampling that overcomes that limitation by employing the localization of the enhanced near-field of a nanometric needle tip. A probe field perturbs the photoemission from the tip, which is induced by a pump pulse, generating a field-dependent current modulation that can easily be captured with our electronic detection scheme. The approach provides reliable characterization of near-petahertz fields. We show that not only the spiral wave-front of visible femtosecond light pulses carrying orbital angular momentum (OAM) can be resolved, but also the field evolution with time in the focal plane. Additionally, our method is polarization sensitive, which makes it applicable to vectorial field reconstruction. 
\end{abstract}

\maketitle

%**************************************************
%**************************************************
\section{Introduction}

The precise knowledge of the electro-magnetic field oscillations of light is not only the backbone of ultrafast science\,\cite{Brabec:2000,Krausz:2009}, but the indispensable prerequisite for many applications such as time-domain terahertz spectroscopy\,\cite{Cocker:2021,Peller:2021,Wimmer:2014} and field-resolved mid-infrared spectroscopy\,\cite{Pupeza:2020,Neuhaus:2022}. %add Refs.
Common techniques for field sampling reaching from the near-infrared to the visible spectral region include attosecond streaking\cite{Hentschel:2001,Itatani:2002,Kienberger:2004,Goulielmakis:2004,Hammond:2016,Wyatt:2016,Kim:2020}, electro-optic sampling \cite{Keiber:2016}, femtosecond streaking\cite{Korobenko:2020}, non-linear photoconductive sampling\cite{Zimin:2021}, or the tunneling ionization with a perturbation for the time-domain observation of an electric field (\textsc{Tiptoe})\,\cite{Park:2018,Saito:2018,Hwang:2019,Cho:2019,Cho:2021,Bionta:2021,Liu:2021,Liu:2022}.
While the sampling of electric field waveforms in the time domain is well established, its simultaneous spatial characterization has remained challenging.  
Whereas for terahertz radiation, where the wavelength is much longer than for visible light, sub-focal size resolution can be achieved, for instance, by using cameras\,\cite{Zhao:2019} or small sensors\,\cite{Mitrofanov:2017}, spatially resolved field measurements in the focus of a visible light beam are challenging as the typical focal size is of the order of a few micrometers. Here, a sub-micrometer probe is necessary.

% thoughts on measurement of OAM fields with conventional methods
Furthermore, measuring the field of an optical vortex beam brings additional difficulties related to the spatial phase structure in the focus.
Because of its helical phase, the field has a \(\pi\) phase-shift at opposite sides of the OAM mode\,\cite{Allen:1992,Yao:2011}. 
Thus, for measurement techniques that are only sensitive to the focal averaged field, a complete cancellation of the signal can be expected. 
Consequently, the field-resolved measurement of vortex beams requires sub-focal spatial sampling. 
In attosecond streaking experiments, an extreme ultraviolet (XUV) pulse is used to sample the electric field of a co-propagating (near-petahertz) field. Here, the XUV-focal size is much smaller than the sampling beam focal size, so the spatio-temporal characterization may be possible. Such measurements, however, require complex vacuum setups, electron spectroscopy, and would be experimentally demanding. It is thus not surprising that the use of attosecond streaking for the spatial reconstruction of near-petahertz vortex fields has not been reported yet. 
The measurement of vortex fields has mostly been limited to the measurement of the corresponding OAM carried by the light\,\cite{Yao:2011,Rego:2019,Fang:2022}, not the field itself.

%needed to move figure up here. Otherwise, it is on the next page 
\begin{figure*}[htbp!] 
    \centering
    \includegraphics[width=180mm]{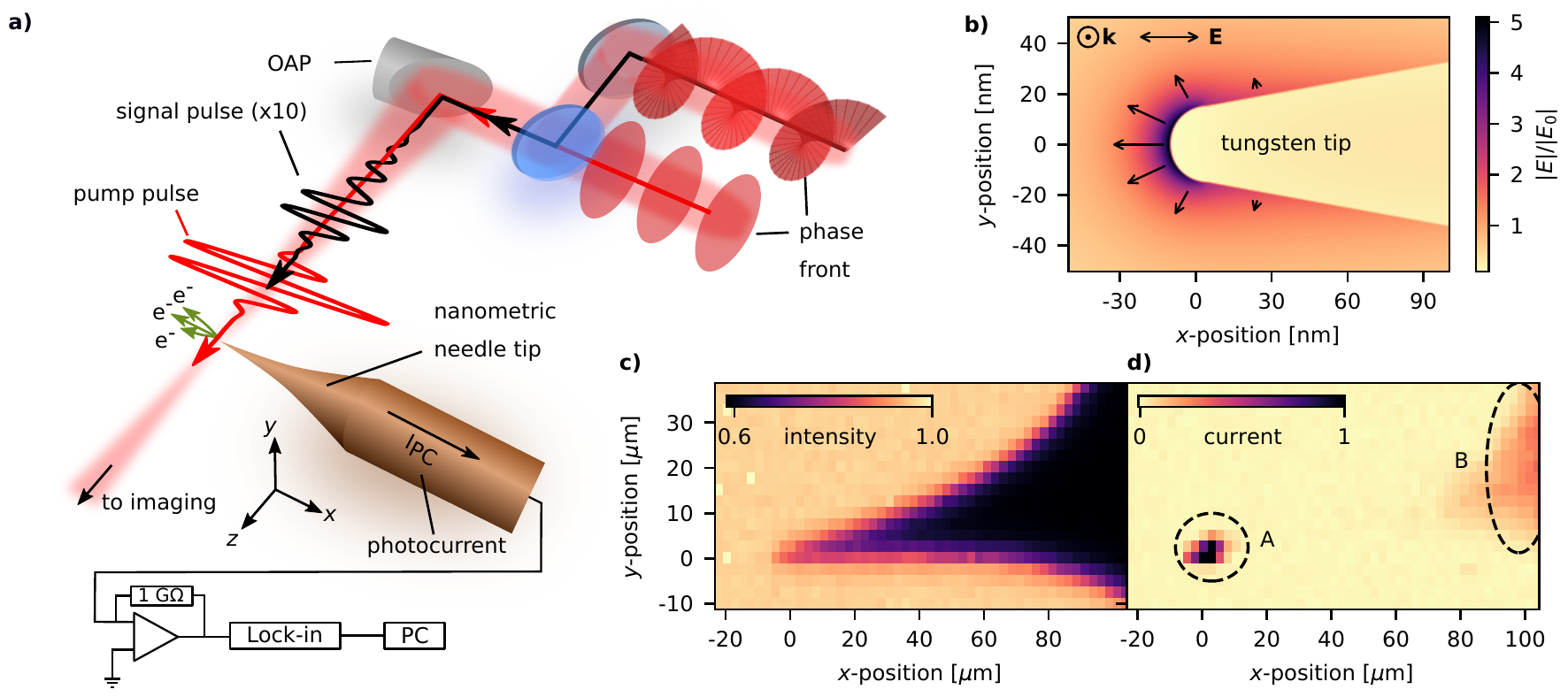}
	\caption[The nano\textsc{Tiptoe} approach for spatio-temporal field sampling]{The nano\textsc{Tiptoe} approach for spatio-temporal field sampling: \textbf{a)} Experimental setup: The pump pulse (red line) and signal pulse (black line) were focused onto a tungsten tip with an off-axis parabola (OAP). The photocurrent generated by the emitted electrons was trans-impedance amplified and lock-in detected. \textbf{b)} Finite-difference time-domain simulations of the field around an exemplary tungsten tip with $10.5^\circ$ half-opening angle and a radius of $r=15$\,nm have shown that the field is enhanced by a factor of around 5, in agreement with literature values\,\cite{Thomas:2015}. The arrows indicate the near-field polarization. \textbf{c)}. Shadowgraphy image obtained by scanning the tip across the $x,y$-plane while detecting the transmitted intensity. The opening angle in the experiment was determined to be \((21\pm2)^\circ\). \textbf{d)} Detected ionization current as a function of position. The signal in region A corresponds to the ionization at the apex, whereas the signal in region B is due to a small contribution from ionization near the rear-end of the needle shank, which can be spatially discriminated.}
	\label{fig:F1}
\end{figure*}
%near field methods
To date, approaches using near-field methods to achieve spatial resolution could either not resolve the electric field of light itself\,\cite{Barwick:2007,Hoff:2017,Garg:2020,Garg:2021}, were limited to much lower frequency ranges in the THz regime\,\cite{Wimmer:2014,Peller:2021}, or detected a spatially averaged signal from several sensors\,\cite{Bionta:2021,Liu:2022}.
We note, that the work presented in Ref.\,\cite{Liu:2022} has demonstrated a measurement of spatio-temporally coupled laser pulses. Here, the frequencies were limited to the infrared range and spatial resolution only achieved in one dimension (limited by the pixel size to (5.2\,\(\mu\)m)).
Here, we overcome these limitations by employing a single nanometric needle tip as a localized probe for the near-field sampling of femtosecond light fields. 
The approach, termed nano\textsc{Tiptoe} and illustrated in Fig.\,\ref{fig:F1}a), inherits the method for sampling the electric field from \textsc{Tiptoe}\,\cite{Park:2018} and achieves high spatial resolution from field localization at the nanometric needle tip.

In nano\textsc{Tiptoe}, a few-cycle pump pulse drives electron emission in the tunneling regime that depends nonlinearly on the electric field. 
Due to this nonlinearity and a short pump pulse, the photoemission is limited to the strongest half-cycle of the laser pulse, and suppressed otherwise. 
We note that only electric field vectors pointing into the surface cause photoemission from the nanometric needle tip\,\cite{Yalunin:2011}. 
Similar to \textsc{Tiptoe}, the emission burst during the strongest half-cycle opens a sub-cycle temporal gate that is perturbed linearly with the signal pulse\,\cite{Park:2018}, enabling characterization of its field. 
We measured the resulting photoemission current from the needle tip after transimpedance amplification employing lock-in-detection. 
Importantly, the fields driving photoemission were the locally enhanced near-fields, which were strongest near the apex of the tip, cf. Fig.\,\ref{fig:F1}b). 
The current measurement approach makes complex ultra-high-vacuum-based time-of-flight spectroscopy\,\cite{Schoetz:2017,Hoff:2017} obsolete, which is a major advance in simplifying such measurements.

\section{Experimental Details}
The output of a commercial 10 kHz Ti:Sa chirped pulse amplifier is broadened in a hollow-core fiber to an octave spanning spectrum ranging from 500\,nm to 1000\,nm, with a central wavelength of 750\,nm. 
The pulses are then compressed to a duration of around 4.2\,fs using chirped mirrors (UFI PC70). 
The laser beam was actively stabilized in angle and position.
The pulses were split into a strong pump pulse and a weak signal pulse in a Mach-Zehnder interferometer (not shown in Fig.\,\ref{fig:F1}a)), where the signal pulse is chopped at half the repetition rate. 
To facilitate the precise control of the delay between signal and pump pulses, the pump arm is provided with a retro-reflector mounted on a closed-loop piezo-stage (MCL OPM100) with 100\,$\mu$m travel range. 
The beams were focused with variable temporal delay onto a nanometric tungsten needle tip inside a vacuum chamber (\(2\times 10^{-3}\)\,mbar) using an off-axis parabolic mirror (OAP, \(f=101.6\)\,mm). 
For the shadow image in Fig.\,\ref{fig:F1}c), we used \(f=25.4\)\,mm to get a sharper contour. 
The needle was directly soldered to a BNC pin, which was mounted onto a 3D closed-loop piezo stick-slip stage.
The photocurrent is amplified by \(10^9\)\,V/A using a low-noise high-gain transimpedance amplifier (FEMTO DLPCA-200) and detected using a lock-in amplifier (Z\"urich Instruments HF2LI).
The upper cut-off frequency (\(f_{-3\,\text{dB}}\)) of the transimpedance amplifier is 1.1\,kHz, which is below the repetition rate of our laser of 10\,kHz and 5\,kHz for pump and signal beam, respectively. 
The expected damping of the signal can be estimated from the measured amplifier response curve provided by the manufacturer to -20\,dBV and -14\,dBV for 10 and 5\,kHz, respectively.
These values correspond to a damping factor of the voltage signal of 10 and 5, in that order.
Whenever we estimated a number of emitted electrons, we took these factors into account. 
For the lock-in detection, we used a demodulation bandwidth of 1.459\,Hz (\(f_{-\text{3\,dB}}\)).
From a noise reference measurement with blocked laser beams, we calculated the normalized noise density to 4.8\,\({\mu\text{V}}/{\sqrt{\text{Hz}}}\) and 4.2\,\({\mu\text{V}}/{\sqrt{\text{Hz}}}\) for the 5- and 10\,kHz components, respectively.
This is close to the specified value of 4.3\,\({\text{fA}}/{\sqrt{\text{Hz}}}\) noise current at our amplification of \(10^9\,{\text V}/{\text A}\).
In order to be detectable, the minimum current modulation thus has to be larger than the noise current times the damping, that is: \(e\cdot n_e/s > 4.8\,\text{fA} \times 5\).
Here, \(e\) is the elementary charge, and \(n_e\) the number of electrons.
This corresponds to a current modulation of at least 30 electrons per shot, assuming a 1\,Hz demodulation bandwidth for illustration purposes.
In the high-gain mode of the amplifier, the amplification bandwidth would be even larger, such that damping becomes negligible, and a modulation of only 6 electrons per shot would become detectable in theory.
We found best signal-to-noise performance, however, in the low-noise mode, where the cut-off frequency was below the repetition rate, as discussed earlier.
The lock-in detection separates the contributions from pump pulse and signal pulse as they have different repetition rates. 
We were therefore able to directly measure a modulation current without the current caused by the pump beam.
For a reference measurement using conventional \textsc{Tiptoe}, a pair of copper electrodes with a distance of \((120\pm15)\,\mu\)m was employed to detect the total ionization yield in gas (i.e., air at 50\,mbar). 
A bias voltage of 10\,V between the electrodes was applied directly by the transimpedance amplifier.
The sampling speed was around 10-14 data points per second, corresponding to a time interval longer than the time-constant of the lock-in amplifier of 47\,ms. 
In addition, the data acquisition was paused for 100\,ms after each step in space, in order to wait for the decay of currents induced by the movement of the tip.

\section{Results and Discussion}

Before performing the actual field measurements, a large raster scan of the nanometric needle tip in the $x,y$-plane in the laser focus was performed to confirm that the current is generated at the apex of the tip with a laser beam polarized along the tip axis (Fig.\,\ref{fig:F1}d)).
Simultaneously, we recorded the transmitted light in an imaging geometry resulting in the shadow image of the tip shown in Fig.\,\ref{fig:F1}c).
The comparison of Figs.\,\ref{fig:F1}c) and d) demonstrates that photoemission occurred predominantly at the apex of the needle tip (region A), whereas currents from sharp features at the needle shank (region B) were prevented by suitable positioning of the tip. 
As there is no emission between region A and B, we conclude that there is no emission from the side of the nanometric needle tip.
Emission away from the tip apex occurs in regions characterized by a large surface roughness. Therefore, the distance between A and B corresponds to the upper limit for the size of the scanning region.
Based on a broad parameter study\,\cite{Thomas:2015} and the experimentally determined enhancement of around \(5.1^{+1.2}_{-0.9}\) (see SI), as well as the opening angle (\((21\pm 2)^\circ\)), we estimated an apex radius of the tip of \(r=14^{+11}_{-7}\)\,nm.

%Figure 2
\begin{figure*}[htp!] 
	\centering\includegraphics[width=180mm]{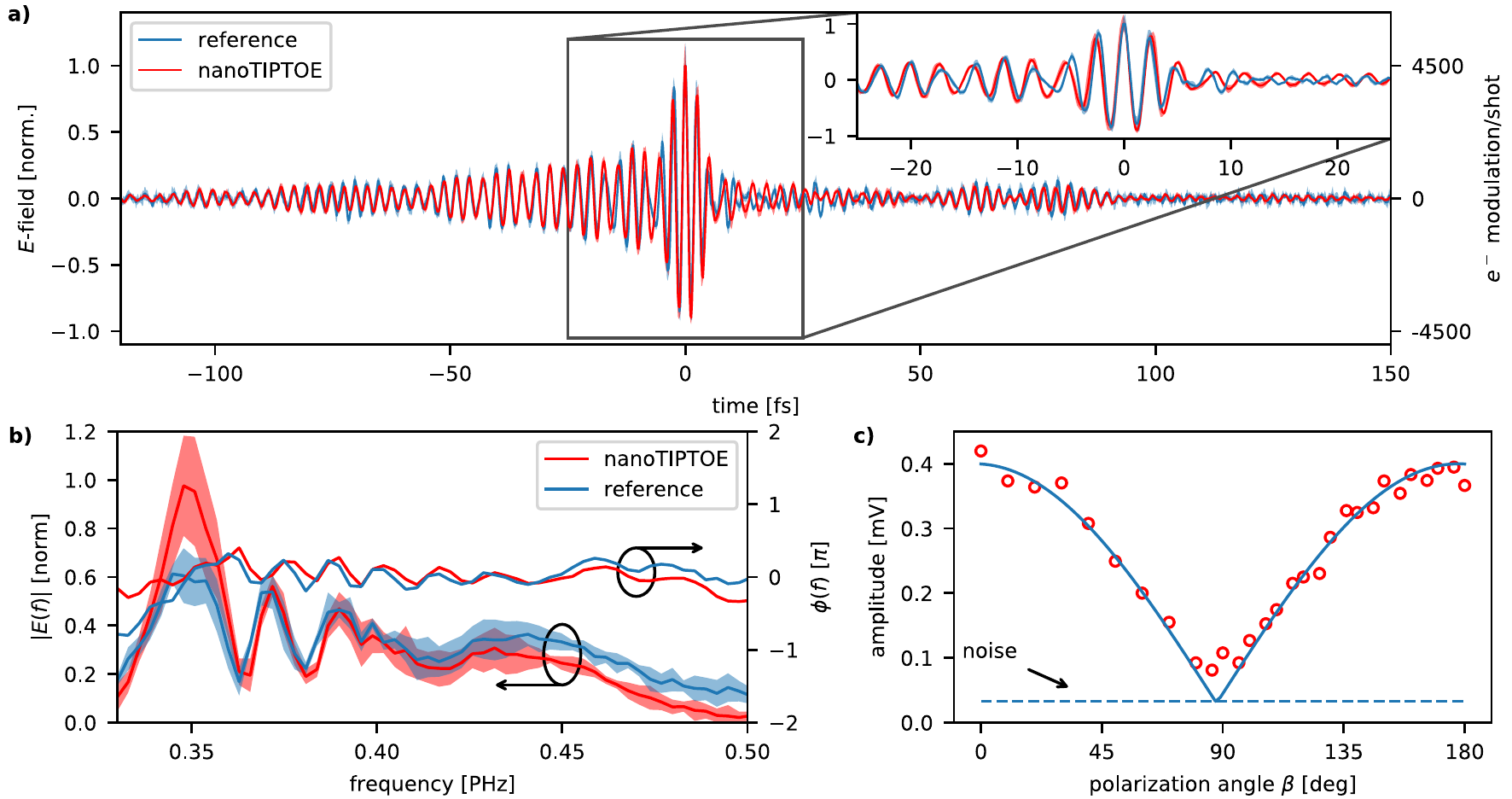}
	\caption[Reference and Polarization]{Field measurement using nano\textsc{Tiptoe}: \textbf{a)} Average and standard deviation of five nano\textsc{Tiptoe} field measurements in the center of the laser focus together with a reference obtained using \textsc{Tiptoe} in gas. We corrected the data for a small offset by removing frequency components below 0.01\,PHz. The enhanced intensities were \((7.2\pm0.9)\times10^{13}\,\frac{\text W}{\text{cm} ^2}\) and \((4.6\pm0.6)\times10^{12}\,\frac{\text W}{\text{cm} ^2}\) for pump and signal beam in the nano\textsc{Tiptoe} regime, respectively. The second axis indicates the current modulation in the nano\textsc{Tiptoe} regime. The excellent agreement of nano\textsc{Tiptoe} and standard \textsc{Tiptoe} can be seen in the inset. \textbf{b)} Spectral amplitudes with error bar and spectral phase, obtained via a Fourier-transform of the data in a). \textbf{c)} Detected amplitudes depending on polarization angle of the linearly-polarized signal beam (red dots). As expected, the signal scales proportional to \(|\cos(\beta)|\). The pump polarization remained the same. }
	\label{fig:F2}
\end{figure*}

The nano\textsc{Tiptoe} measurement obtained for linearly polarized sample and pump pulses (polarized along the needle direction) with the needle tip placed in the center of the focus is presented in Fig.\,\ref{fig:F2}a). 
The obtained waveform is compared to a reference obtained via conventional \textsc{Tiptoe}. 
We note that the enhanced field on the needle tip exhibits a phase shift of typically \(0.2\pi\) to \(0.5\pi\) compared to the incident field\,\cite{Thomas:2015}. 
However, since both beams in nano\textsc{Tiptoe} experience the same shift due to the enhancement, the overall phase difference is zero and therefore does not affect the measurement.
The excellent agreement between the nano\textsc{Tiptoe} measurement and the reference indicates a rather flat spectral response and demonstrates the capability of nano\textsc{Tiptoe} to sample near-petahertz laser fields. 
This conclusion is further supported by the similarity of the measured spectral phases obtained using both methods (see Fig.\,\ref{fig:F2}b)).
The nano\textsc{Tiptoe} measurements are only slightly red-shifted, as evident from the spectral amplitudes of the measured pulses in Fig.\,\ref{fig:F2}b) and the calculated response, see SI Fig.\,S5. 
This difference relates to the response function of the nanometric needle tip. 
The good agreement of the time-domain waveforms indicates a secondary importance of this small red-shift to most applications.

%%%%  nonlinearities and Polarization Scan
To further validate the nano\textsc{Tiptoe} technique for field sampling, we also performed scans of the dispersion of the signal pulse, its carrier-envelope-phase, the field-strength ratio (see corresponding SI sections) and its polarization.
For the investigation of the polarization dependence, we kept the pump beam polarized along the tip axis and rotated the polarization of the signal beam, see Fig.\,\ref{fig:F2}c).
As the superposition of both beams drives the ionization process, we would expect a scaling of the signal in free-space as \(\sim |\cos(\beta)|\), where \(\beta\) is the angle between the polarizations of the two laser pulses (solid blue line in Fig.\,\ref{fig:F2}c)).
However, at a nanostructure, the pump beam generates surface normal near-fields that the signal beam can interfere with (cf. Fig.\,\ref{fig:F1}c) and Ref.\,\cite{Thomas:2015}). 
Our polarization scan (Fig.\,\ref{fig:F2}c)) suggests that the interference of the signal beam with surface normal near-fields has only a minor influence on the signal taken with nano\textsc{Tiptoe}, as the signal amplitude for perpendicular polarization nearly reaches the electronic noise amplitude (dashed blue line, Fig.\,\ref{fig:F2}c)).
Therefore, nano\textsc{Tiptoe} exhibits a polarization sensitivity that allows to map the two-dimensional polarization state.

%% Figure 3
\begin{figure*}[htbp!]
	\centering
	\includegraphics[width=180mm]{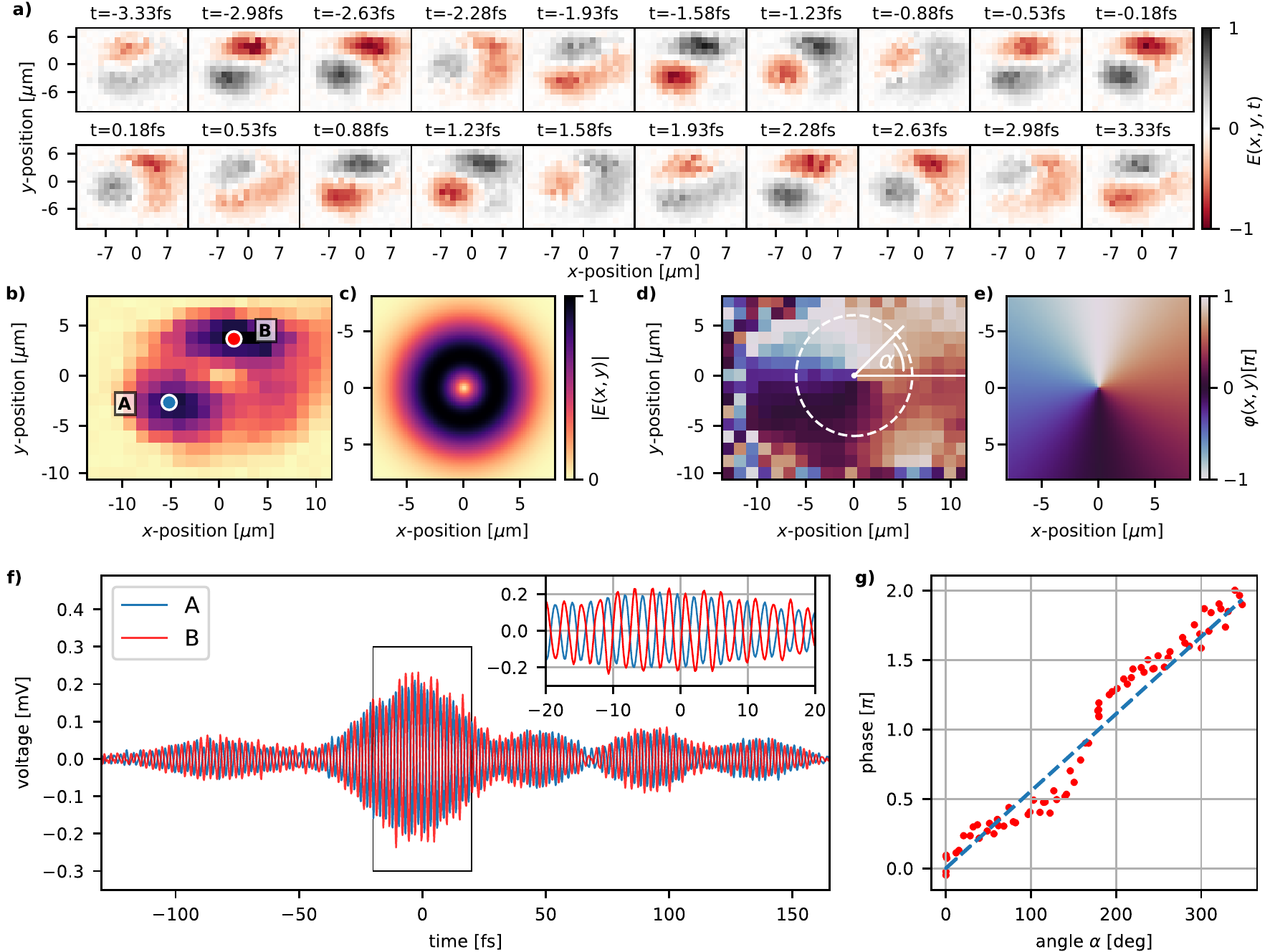}
	\caption[Spatially-resolved Measurements of OAM Beams]{Spatially-resolved measurements of OAM beams: 
	\textbf{a)} The field amplitudes as a function of space and time, normalized to unity. \textbf{b)} The amplitude of the measured current modulation induced by the OAM beam together with the expected shape for the theoretical Laguerre-Gaussian beam shown in \textbf{c)}. 
	Deviations between experimental data and the simple OAM profile can be attributed to astigmatism in the focusing optics as well as to a non-Gaussian mode shape that has been used for the vortex generation. 
	\textbf{d)}
	The extracted phase at each point in the sampled plane exhibits a helical profile as theoretically shown in \textbf{e)}. 
	The angle \(\alpha\) defines a polar coordinate system.
	\textbf{f)} Field sampling with nano\textsc{Tiptoe} at the points A and B at opposite sides of the focus, indicated in b). 
	\textbf{g)} The phase for each point within the dashed circle in d) (red dots) shows a linear trend with \(\alpha\) (dashed blue line).
	The raw data used for the plots in a), b), d) and g) can be seen in supplementary figure\,S1. To generate the plot in f), a Fourier filter from 0.1 to 1\,PHz has been applied in order to remove nonlinear distortions.
	}\label{fig:F3}
\end{figure*}

%raster scan 
Having established that nano\textsc{Tiptoe} provides the electric field waveform of the sampling field, we can now investigate how scanning the needle tip accross the focal plane provides spatially-resolved data. 
Some care has to be taken in such scanning measurements. 
As the measured waveform samples only the relative phase between signal and pump pulse\,\cite{Bionta:2021,Liu:2021}, the mode size of the pump beam has been made smaller by a factor of roughly \(2.4\), causing a larger focal spot size. 
This increase in size  leads to a rather flat pump beam intensity and phase profile over the area that is scanned for the sampling pulse.
In order to demonstrate the capability of nano\textsc{Tiptoe} in spatio-temporal field sampling of near-petahertz fields, we sampled a light beam carrying orbital angular momentum (OAM)\,\cite{Allen:1992,Yao:2011}. The signal beam was shaped by a vortex plate (Vortex Photonics V-780-20-1), that preserves linear polarization, into an OAM beam. 
As the wave plate had a limited bandwidth, a suitable bandpass filter was added, which increased the pulse duration to 33\,fs. 
The pump pulse, however, was not modified such that the temporal gate remained short. 
The vortex beam resulting from the beam shaping of the signal pulse is expected to exhibit a field distribution with a singularity on the propagation axis as well as a helical phase shape (cf. Fig.\,\ref{fig:F1}a) and Fig.\,\ref{fig:F3}c)).

In order to map the evolution of the vortex field in the focal plane, we scanned the tip through the beam while varying the delay over a few oscillations of the most intense part of the signal pulse. The result is depicted in Fig.\,\ref{fig:F3}a). A clear rotational motion of the field amplitudes around the center of the focal spot is observable.
The extracted amplitude and phase as a function of the needle tip position are shown in Fig.\,\ref{fig:F3}b) and d), together with a theoretical expectation for a Laguerre-Gaussian mode (c) and e)), see SI for details. 
An animated version of Fig.\,\ref{fig:F3}a) can be found in the supplementary material.
The field distribution exhibits a typical doughnut-shape, the minimum of which is visible in the middle despite the pump beam being maximal there.
In order to validate that nano\textsc{Tiptoe} provides full spectral resolution at every point in space, we also performed scans over the full pulse length, but only at selected points marked as A and B in Fig.\,\ref{fig:F3}b).
The corresponding data was Fourier-transform filtered and is shown in Fig.\,\ref{fig:F3}f). 
The points A and B were chosen at opposite sides of the mode, since this is where we expect the spatial phase difference to be maximum. Indeed, the corresponding waveforms exhibit a clear \(\pi\)-phase difference, which is due to the OAM of the signal beam.
As can be seen in Fig.\,\ref{fig:F3}d), one of the main features of light carrying orbital angular momentum, the helical phase front, agrees well with the theoretical prediction, Fig.\,\ref{fig:F3}e).
To quantify the spiral phase, we introduced a polar angle \(\alpha\), and evaluated all phase points within the dashed circle in Fig.\,\ref{fig:F3}d).
We fixed the central point and calculated the corresponding \(\alpha\) for every data point in the region of interest. The result (red dots in Fig.\,\ref{fig:F3}\,g)), is in qualitative agreement with the expected linear increase of the phase with \(\alpha\). 
We attribute the small deviations from the linear scaling for \(\alpha \in [120^\circ:220^\circ]\) to a curved pulse front of the pump beam, which is even visible without vortex plate, see SI for details. 

\section{Conclusions}
In conclusion, we have demonstrated that nano\textsc{Tiptoe} enables the spatially resolved measurement of near-petahertz optical field oscillations with sub-cycle resolution.
The localized probe enabled the spatio-temporal characterization of optical fields by employing a nanometric needle tip instead of the conventional electrodes.
The field enhancement allowed the characterization of laser fields with moderate intensity - a major advance compared to techniques requiring high-power laser sources.
As compared to previous approaches for electronic field detection with nanotips, which were often limited to the low terahertz region\,\cite{Wimmer:2014,Peller:2021}, nano\textsc{Tiptoe} increases the temporal resolution by nearly three orders of magnitude.
While the detected bandwidth is comparable to that in Ref.\,\cite{Schoetz:2017} using attosecond streaking spectroscopy, the nano\textsc{Tiptoe} approach is much simpler and avoids a complex vacuum beamline.
A combination of nano\textsc{Tiptoe} with latest approaches in time-resolved scanning tunneling microscopy\,\cite{Garg:2020,Garg:2021} seems promising in characterizing the light-induced near-fields of a nanometric sample with attosecond precision. 
Theoretically, the resolution here is limited by the interaction of the nanometric needle with the field close to the apex, as extensively studied in Ref.\,\cite{Thomas:2015}. 
For our tip geometry, we expect a maximum resolution in the order of the tip diameter, and even down to 1\,nm\cite{Thomas:2013} using smaller tips.
Finally, orienting the needle along the propagation direction of the laser beam\,\cite{Bouhelier:2003} may pave the way towards the measurement of the longitudinal component of strongly focused light with nano\textsc{Tiptoe}. Such measurements would offer a more detailed understanding of the properties of focused light in superresolution microscopy.

\section*{Author Contributions}
J.B. and J.S. contributed equally to this work. J.S. and M.F.K. conceived the nano\textsc{Tiptoe} concept. 
P.H. contributed expertise on nanometric needle tips. 
A.S., M.F.K., and P.C. conceived the experiment with OAM beams. 
J.B., J.S., A.M., and N.S. performed the measurements. 
Z.W. and P.R. supported laser operations. 
The data was analyzed by J.B. and J.S. 
The manuscript was written by J.B., J.S., B.B. and M.F.K. and reviewed by all authors.

\section*{Funding} Deutsche Forschungsgemeinschaft (DFG) (SPP1840 (QUTIF),LMUexcellent); European ResearchCouncil (ERC) (FETopen PetaCOM, FETlaunchpad FIELDTECH); Alexander von Humboldt Stiftung; Max-Planck Gesellschaft (MPG) (IMPRS-APS, MPSP, Fellow Program); ERC Adv. Gr. AccelOnChip; U.S. Department of Energy, Office of Science, Basic Energy Sciences, Scientific User Facilities Division (DE-AC02-76SF00515); US Defense Threat Reduction Agency (DTRA) (HDTRA1-19-1-0026); University of Ottawa; NRC Joint Centre for Extreme Photonics; Natural Sciences and Engineering Research Council of Canada (NSERC). 

\section*{Acknowledgments} We are grateful for support by Ferenc Krausz providing suitable laboratories and for support by Thomas Nubbemeyer and Maximilian Seeger in laser operations. We acknowledge fruitful discussions with Matthew Weidman, Vladislav Yakovlev, Nicholas Karpowicz, and Ferenc Krausz. The experiments were carried out at LMU and MPQ. J.S., and A.M. acknowledge support by the Max Planck Society via the IMPRS for Advanced Photon Science. J.B. acknowledges support by the Max Planck School of Photonics. M.F.K. acknowledges support by the Max Planck Society via the Max Planck Fellow program. M.F.K.'s work at SLAC is supported by the U.S. Department of Energy, Office of Science, Basic Energy Sciences, Scientific User Facilities Division. P.B.C. acknowledges funds from the US Defense Threat Reduction Agency (DTRA) and the University of Ottawa. P.B.C. and A.S. acknowledge support from NRC Joint Centre for Extreme Photonics and also from the Natural Sciences and Engineering Research Council of Canada (NSERC).

\section*{Disclosures} The authors declare no competing interests.

\section*{Data availability} Data underlying the results presented in this paper are not publicly available at this time but may be obtained from the authors upon reasonable request.

\section*{Supplemental document}
See Supplement 1 for supporting content as well as the supplementary animation.


\begin{thebibliography}{10}
\expandafter\ifx\csname url\endcsname\relax
  \def\url#1{\texttt{#1}}\fi
\expandafter\ifx\csname urlprefix\endcsname\relax\def\urlprefix{URL }\fi
\providecommand{\bibinfo}[2]{#2}
\providecommand{\eprint}[2][]{\url{#2}}

\bibitem{Brabec:2000}
\bibinfo{author}{Brabec, T.} \& \bibinfo{author}{Krausz, F.}
\newblock \bibinfo{title}{Intense few-cycle laser fields: Frontiers of
  nonlinear optics}.
\newblock \emph{\bibinfo{journal}{Rev. Mod. Phys.}}
  \textbf{\bibinfo{volume}{72}}, \bibinfo{pages}{545--591}
  (\bibinfo{year}{2000}).
\newblock \urlprefix\url{https://link.aps.org/doi/10.1103/RevModPhys.72.545}.

\bibitem{Krausz:2009}
\bibinfo{author}{Krausz, F.} \& \bibinfo{author}{Ivanov, M.}
\newblock \bibinfo{title}{Attosecond physics}.
\newblock \emph{\bibinfo{journal}{Rev. Mod. Phys.}}
  \textbf{\bibinfo{volume}{81}}, \bibinfo{pages}{163--234}
  (\bibinfo{year}{2009}).
\newblock \urlprefix\url{https://link.aps.org/doi/10.1103/RevModPhys.81.163}.

\bibitem{Cocker:2021}
\bibinfo{author}{Cocker, T.~L.}, \bibinfo{author}{Jelic, V.},
  \bibinfo{author}{Hillenbrand, R.} \& \bibinfo{author}{Hegmann, F.~A.}
\newblock \bibinfo{title}{Nanoscale terahertz scanning probe microscopy}.
\newblock \emph{\bibinfo{journal}{Nature Photonics}}
  \textbf{\bibinfo{volume}{15}}, \bibinfo{pages}{558--569}
  (\bibinfo{year}{2021}).
\newblock \urlprefix\url{https://doi.org/10.1038/s41566-021-00835-6}.

\bibitem{Peller:2021}
\bibinfo{author}{Peller, D.} \emph{et~al.}
\newblock \bibinfo{title}{Quantitative sampling of atomic-scale electromagnetic
  waveforms}.
\newblock \emph{\bibinfo{journal}{Nature Photonics}}
  \textbf{\bibinfo{volume}{15}}, \bibinfo{pages}{143--147}
  (\bibinfo{year}{2021}).
\newblock \urlprefix\url{https://doi.org/10.1038/s41566-020-00720-8}.

\bibitem{Wimmer:2014}
\bibinfo{author}{Wimmer, L.} \emph{et~al.}
\newblock \bibinfo{title}{Terahertz control of nanotip photoemission}.
\newblock \emph{\bibinfo{journal}{Nature Physics}}
  \textbf{\bibinfo{volume}{10}}, \bibinfo{pages}{432--436}
  (\bibinfo{year}{2014}).
\newblock \urlprefix\url{https://doi.org/10.1038/nphys2974}.

\bibitem{Pupeza:2020}
\bibinfo{author}{Pupeza, I.} \emph{et~al.}
\newblock \bibinfo{title}{Field-resolved infrared spectroscopy of biological
  systems}.
\newblock \emph{\bibinfo{journal}{Nature}} \textbf{\bibinfo{volume}{577}},
  \bibinfo{pages}{52--59} (\bibinfo{year}{2020}).
\newblock \urlprefix\url{https://doi.org/10.1038/s41586-019-1850-7}.

\bibitem{Neuhaus:2022}
\bibinfo{author}{Neuhaus, M.} \emph{et~al.}
\newblock \bibinfo{title}{Transient field-resolved reflectometry at 50--100
  thz}.
\newblock \emph{\bibinfo{journal}{Optica}} \textbf{\bibinfo{volume}{9}},
  \bibinfo{pages}{42--49} (\bibinfo{year}{2022}).
\newblock
  \urlprefix\url{http://opg.optica.org/optica/abstract.cfm?URI=optica-9-1-42}.

\bibitem{Hentschel:2001}
\bibinfo{author}{Hentschel, M.} \emph{et~al.}
\newblock \bibinfo{title}{Attosecond metrology}.
\newblock \emph{\bibinfo{journal}{Nature}} \textbf{\bibinfo{volume}{414}},
  \bibinfo{pages}{509--513} (\bibinfo{year}{2001}).
\newblock \urlprefix\url{https://doi.org/10.1038/35107000}.

\bibitem{Itatani:2002}
\bibinfo{author}{Itatani, J.} \emph{et~al.}
\newblock \bibinfo{title}{Attosecond streak camera}.
\newblock \emph{\bibinfo{journal}{Phys. Rev. Lett.}}
  \textbf{\bibinfo{volume}{88}}, \bibinfo{pages}{173903}
  (\bibinfo{year}{2002}).
\newblock
  \urlprefix\url{https://link.aps.org/doi/10.1103/PhysRevLett.88.173903}.

\bibitem{Kienberger:2004}
\bibinfo{author}{Kienberger, R.} \emph{et~al.}
\newblock \bibinfo{title}{Atomic transient recorder}.
\newblock \emph{\bibinfo{journal}{Nature}} \textbf{\bibinfo{volume}{427}},
  \bibinfo{pages}{817--821} (\bibinfo{year}{2004}).
\newblock \urlprefix\url{https://doi.org/10.1038/nature02277}.

\bibitem{Goulielmakis:2004}
\bibinfo{author}{Goulielmakis, E.} \emph{et~al.}
\newblock \bibinfo{title}{Direct measurement of light waves}.
\newblock \emph{\bibinfo{journal}{Science}} \textbf{\bibinfo{volume}{305}},
  \bibinfo{pages}{1267--1269} (\bibinfo{year}{2004}).
\newblock \urlprefix\url{https://science.sciencemag.org/content/305/5688/1267}.
\newblock
  \eprint{https://science.sciencemag.org/content/305/5688/1267.full.pdf}.

\bibitem{Hammond:2016}
\bibinfo{author}{Hammond, T.~J.}, \bibinfo{author}{Brown, G.~G.},
  \bibinfo{author}{Kim, K.~T.}, \bibinfo{author}{Villeneuve, D.~M.} \&
  \bibinfo{author}{Corkum, P.~B.}
\newblock \bibinfo{title}{Attosecond pulses measured from the attosecond
  lighthouse}.
\newblock \emph{\bibinfo{journal}{Nature Photonics}}
  \textbf{\bibinfo{volume}{10}}, \bibinfo{pages}{171--175}
  (\bibinfo{year}{2016}).
\newblock \urlprefix\url{https://doi.org/10.1038/nphoton.2015.271}.

\bibitem{Wyatt:2016}
\bibinfo{author}{Wyatt, A.~S.} \emph{et~al.}
\newblock \bibinfo{title}{Attosecond sampling of arbitrary optical waveforms}.
\newblock \emph{\bibinfo{journal}{Optica}} \textbf{\bibinfo{volume}{3}},
  \bibinfo{pages}{303--310} (\bibinfo{year}{2016}).
\newblock
  \urlprefix\url{http://www.osapublishing.org/optica/abstract.cfm?URI=optica-3-3-303}.

\bibitem{Kim:2020}
\bibinfo{author}{Kim, Y.~H.} \emph{et~al.}
\newblock \bibinfo{title}{Attosecond streaking using a rescattered electron in
  an intense laser field}.
\newblock \emph{\bibinfo{journal}{Scientific Reports}}
  \textbf{\bibinfo{volume}{10}}, \bibinfo{pages}{22075} (\bibinfo{year}{2020}).
\newblock \urlprefix\url{https://doi.org/10.1038/s41598-020-79034-2}.

\bibitem{Keiber:2016}
\bibinfo{author}{Keiber, S.} \emph{et~al.}
\newblock \bibinfo{title}{Electro-optic sampling of near-infrared waveforms}.
\newblock \emph{\bibinfo{journal}{Nature Photonics}}
  \textbf{\bibinfo{volume}{10}}, \bibinfo{pages}{159--162}
  (\bibinfo{year}{2016}).
\newblock \urlprefix\url{https://doi.org/10.1038/nphoton.2015.269}.

\bibitem{Korobenko:2020}
\bibinfo{author}{Korobenko, A.} \emph{et~al.}
\newblock \bibinfo{title}{Femtosecond streaking in ambient air}.
\newblock \emph{\bibinfo{journal}{Optica}} \textbf{\bibinfo{volume}{7}},
  \bibinfo{pages}{1372--1376} (\bibinfo{year}{2020}).
\newblock
  \urlprefix\url{http://www.osapublishing.org/optica/abstract.cfm?URI=optica-7-10-1372}.

\bibitem{Zimin:2021}
\bibinfo{author}{Zimin, D.} \emph{et~al.}
\newblock \bibinfo{title}{Petahertz-scale nonlinear photoconductive sampling in
  air}.
\newblock \emph{\bibinfo{journal}{Optica}} \textbf{\bibinfo{volume}{8}},
  \bibinfo{pages}{586--590} (\bibinfo{year}{2021}).
\newblock
  \urlprefix\url{http://www.osapublishing.org/optica/abstract.cfm?URI=optica-8-5-586}.

\bibitem{Park:2018}
\bibinfo{author}{Park, S.~B.} \emph{et~al.}
\newblock \bibinfo{title}{Direct sampling of a light wave in air}.
\newblock \emph{\bibinfo{journal}{Optica}} \textbf{\bibinfo{volume}{5}},
  \bibinfo{pages}{402--408} (\bibinfo{year}{2018}).
\newblock
  \urlprefix\url{http://www.osapublishing.org/optica/abstract.cfm?URI=optica-5-4-402}.

\bibitem{Saito:2018}
\bibinfo{author}{Saito, N.}, \bibinfo{author}{Ishii, N.},
  \bibinfo{author}{Kanai, T.} \& \bibinfo{author}{Itatani, J.}
\newblock \bibinfo{title}{All-optical characterization of the two-dimensional
  waveform and the gouy phase of an infrared pulse based on plasma fluorescence
  of gas}.
\newblock \emph{\bibinfo{journal}{Opt. Express}} \textbf{\bibinfo{volume}{26}},
  \bibinfo{pages}{24591--24601} (\bibinfo{year}{2018}).
\newblock
  \urlprefix\url{http://www.opticsexpress.org/abstract.cfm?URI=oe-26-19-24591}.

\bibitem{Hwang:2019}
\bibinfo{author}{Hwang, S.~I.} \emph{et~al.}
\newblock \bibinfo{title}{Generation of a single-cycle pulse using a two-stage
  compressor and its temporal characterization using a tunnelling ionization
  method}.
\newblock \emph{\bibinfo{journal}{Scientific Reports}}
  \textbf{\bibinfo{volume}{9}}, \bibinfo{pages}{1613} (\bibinfo{year}{2019}).
\newblock \urlprefix\url{https://doi.org/10.1038/s41598-018-38220-z}.

\bibitem{Cho:2019}
\bibinfo{author}{Cho, W.} \emph{et~al.}
\newblock \bibinfo{title}{Temporal characterization of femtosecond laser pulses
  using tunneling ionization in the uv, visible, and mid-ir ranges}.
\newblock \emph{\bibinfo{journal}{Scientific Reports}}
  \textbf{\bibinfo{volume}{9}}, \bibinfo{pages}{16067} (\bibinfo{year}{2019}).
\newblock \urlprefix\url{https://doi.org/10.1038/s41598-019-52237-y}.

\bibitem{Cho:2021}
\bibinfo{author}{Cho, W.}, \bibinfo{author}{Shin, J.-u.} \&
  \bibinfo{author}{Kim, K.~T.}
\newblock \bibinfo{title}{Reconstruction algorithm for tunneling ionization
  with a perturbation for the time-domain observation of an electric-field}.
\newblock \emph{\bibinfo{journal}{Scientific Reports}}
  \textbf{\bibinfo{volume}{11}}, \bibinfo{pages}{13014} (\bibinfo{year}{2021}).
\newblock \urlprefix\url{https://doi.org/10.1038/s41598-021-92454-y}.

\bibitem{Bionta:2021}
\bibinfo{author}{Bionta, M.~R.} \emph{et~al.}
\newblock \bibinfo{title}{On-chip sampling of optical fields with attosecond
  resolution}.
\newblock \emph{\bibinfo{journal}{Nature Photonics}}
  \textbf{\bibinfo{volume}{15}}, \bibinfo{pages}{456--460}
  (\bibinfo{year}{2021}).
\newblock \urlprefix\url{https://doi.org/10.1038/s41566-021-00792-0}.

\bibitem{Liu:2021}
\bibinfo{author}{Liu, Y.} \emph{et~al.}
\newblock \bibinfo{title}{All-optical sampling of few-cycle infrared pulses
  using tunneling in a solid}.
\newblock \emph{\bibinfo{journal}{Photon. Res.}} \textbf{\bibinfo{volume}{9}},
  \bibinfo{pages}{929--936} (\bibinfo{year}{2021}).
\newblock
  \urlprefix\url{http://www.osapublishing.org/prj/abstract.cfm?URI=prj-9-6-929}.

\bibitem{Liu:2022}
\bibinfo{author}{Liu, Y.}, \bibinfo{author}{Beetar, J.~E.},
  \bibinfo{author}{Nesper, J.}, \bibinfo{author}{Gholam-Mirzaei, S.} \&
  \bibinfo{author}{Chini, M.}
\newblock \bibinfo{title}{Single-shot measurement of few-cycle optical
  waveforms on a chip}.
\newblock \emph{\bibinfo{journal}{Nature Photonics}}
  \textbf{\bibinfo{volume}{16}}, \bibinfo{pages}{109--112}
  (\bibinfo{year}{2022}).
\newblock \urlprefix\url{https://doi.org/10.1038/s41566-021-00924-6}.

\bibitem{Zhao:2019}
\bibinfo{author}{Zhao, J.}, \bibinfo{author}{E, Y.}, \bibinfo{author}{Williams,
  K.}, \bibinfo{author}{Zhang, X.-C.} \& \bibinfo{author}{Boyd, R.~W.}
\newblock \bibinfo{title}{Spatial sampling of terahertz fields with
  sub-wavelength accuracy via probe-beam encoding}.
\newblock \emph{\bibinfo{journal}{Light: Science {\&} Applications}}
  \textbf{\bibinfo{volume}{8}}, \bibinfo{pages}{55} (\bibinfo{year}{2019}).
\newblock \urlprefix\url{https://doi.org/10.1038/s41377-019-0166-6}.

\bibitem{Mitrofanov:2017}
\bibinfo{author}{Mitrofanov, O.} \emph{et~al.}
\newblock \bibinfo{title}{Near-field terahertz probes with room-temperature
  nanodetectors for subwavelength resolution imaging}.
\newblock \emph{\bibinfo{journal}{Scientific Reports}}
  \textbf{\bibinfo{volume}{7}}, \bibinfo{pages}{44240} (\bibinfo{year}{2017}).
\newblock \urlprefix\url{https://doi.org/10.1038/srep44240}.

\bibitem{Allen:1992}
\bibinfo{author}{Allen, L.}, \bibinfo{author}{Beijersbergen, M.~W.},
  \bibinfo{author}{Spreeuw, R. J.~C.} \& \bibinfo{author}{Woerdman, J.~P.}
\newblock \bibinfo{title}{Orbital angular momentum of light and the
  transformation of laguerre-gaussian laser modes}.
\newblock \emph{\bibinfo{journal}{Phys. Rev. A}} \textbf{\bibinfo{volume}{45}},
  \bibinfo{pages}{8185--8189} (\bibinfo{year}{1992}).
\newblock \urlprefix\url{https://link.aps.org/doi/10.1103/PhysRevA.45.8185}.

\bibitem{Yao:2011}
\bibinfo{author}{Yao, A.~M.} \& \bibinfo{author}{Padgett, M.~J.}
\newblock \bibinfo{title}{Orbital angular momentum: origins, behavior and
  applications}.
\newblock \emph{\bibinfo{journal}{Adv. Opt. Photon.}}
  \textbf{\bibinfo{volume}{3}}, \bibinfo{pages}{161--204}
  (\bibinfo{year}{2011}).
\newblock \urlprefix\url{http://aop.osa.org/abstract.cfm?URI=aop-3-2-161}.

\bibitem{Rego:2019}
\bibinfo{author}{Rego, L.} \emph{et~al.}
\newblock \bibinfo{title}{Generation of extreme-ultraviolet beams with
  time-varying orbital angular momentum}.
\newblock \emph{\bibinfo{journal}{Science}} \textbf{\bibinfo{volume}{364}},
  \bibinfo{pages}{eaaw9486} (\bibinfo{year}{2019}).
\newblock
  \urlprefix\url{https://www.science.org/doi/abs/10.1126/science.aaw9486}.
\newblock \eprint{https://www.science.org/doi/pdf/10.1126/science.aaw9486}.

\bibitem{Fang:2022}
\bibinfo{author}{Fang, Y.} \emph{et~al.}
\newblock \bibinfo{title}{Probing the orbital angular momentum of intense
  vortex pulses with strong-field ionization}.
\newblock \emph{\bibinfo{journal}{Light: Science {\&} Applications}}
  \textbf{\bibinfo{volume}{11}}, \bibinfo{pages}{34} (\bibinfo{year}{2022}).
\newblock \urlprefix\url{https://doi.org/10.1038/s41377-022-00726-7}.

\bibitem{Thomas:2015}
\bibinfo{author}{Thomas, S.}, \bibinfo{author}{Wachter, G.},
  \bibinfo{author}{Lemell, C.}, \bibinfo{author}{Burgdörfer, J.} \&
  \bibinfo{author}{Hommelhoff, P.}
\newblock \bibinfo{title}{Large optical field enhancement for nanotips with
  large opening angles}.
\newblock \emph{\bibinfo{journal}{New Journal of Physics}}
  \textbf{\bibinfo{volume}{17}}, \bibinfo{pages}{063010}
  (\bibinfo{year}{2015}).
\newblock \urlprefix\url{https://doi.org/10.1088/1367-2630/17/6/063010}.

\bibitem{Barwick:2007}
\bibinfo{author}{Barwick, B.} \emph{et~al.}
\newblock \bibinfo{title}{Laser-induced ultrafast electron emission from a
  field emission tip}.
\newblock \emph{\bibinfo{journal}{New Journal of Physics}}
  \textbf{\bibinfo{volume}{9}}, \bibinfo{pages}{142--142}
  (\bibinfo{year}{2007}).
\newblock \urlprefix\url{https://doi.org/10.1088/1367-2630/9/5/142}.

\bibitem{Hoff:2017}
\bibinfo{author}{Hoff, D.} \emph{et~al.}
\newblock \bibinfo{title}{Tracing the phase of focused broadband laser pulses}.
\newblock \emph{\bibinfo{journal}{Nature Physics}}
  \textbf{\bibinfo{volume}{13}}, \bibinfo{pages}{947--951}
  (\bibinfo{year}{2017}).
\newblock \urlprefix\url{https://doi.org/10.1038/nphys4185}.

\bibitem{Garg:2020}
\bibinfo{author}{Garg, M.} \& \bibinfo{author}{Kern, K.}
\newblock \bibinfo{title}{Attosecond coherent manipulation of electrons in
  tunneling microscopy}.
\newblock \emph{\bibinfo{journal}{Science}} \textbf{\bibinfo{volume}{367}},
  \bibinfo{pages}{411--415} (\bibinfo{year}{2020}).
\newblock \urlprefix\url{https://science.sciencemag.org/content/367/6476/411}.
\newblock
  \eprint{https://science.sciencemag.org/content/367/6476/411.full.pdf}.

\bibitem{Garg:2021}
\bibinfo{author}{Garg, M.} \emph{et~al.}
\newblock \bibinfo{title}{Real-space subfemtosecond imaging of quantum
  electronic coherences in molecules}.
\newblock \emph{\bibinfo{journal}{Nature Photonics}}  (\bibinfo{year}{2021}).
\newblock \urlprefix\url{https://doi.org/10.1038/s41566-021-00929-1}.

\bibitem{Yalunin:2011}
\bibinfo{author}{Yalunin, S.~V.}, \bibinfo{author}{Gulde, M.} \&
  \bibinfo{author}{Ropers, C.}
\newblock \bibinfo{title}{Strong-field photoemission from surfaces: Theoretical
  approaches}.
\newblock \emph{\bibinfo{journal}{Phys. Rev. B}} \textbf{\bibinfo{volume}{84}},
  \bibinfo{pages}{195426} (\bibinfo{year}{2011}).
\newblock \urlprefix\url{https://link.aps.org/doi/10.1103/PhysRevB.84.195426}.

\bibitem{Schoetz:2017}
\bibinfo{author}{Schötz, J.} \emph{et~al.}
\newblock \bibinfo{title}{Reconstruction of nanoscale near fields by attosecond
  streaking}.
\newblock \emph{\bibinfo{journal}{IEEE Journal of Selected Topics in Quantum
  Electronics}} \textbf{\bibinfo{volume}{23}}, \bibinfo{pages}{77--87}
  (\bibinfo{year}{2017}).

\bibitem{Thomas:2013}
\bibinfo{author}{Thomas, S.}, \bibinfo{author}{Krüger, M.},
  \bibinfo{author}{Förster, M.}, \bibinfo{author}{Schenk, M.} \&
  \bibinfo{author}{Hommelhoff, P.}
\newblock \bibinfo{title}{Probing of optical near-fields by electron
  rescattering on the 1 nm scale}.
\newblock \emph{\bibinfo{journal}{Nano Letters}} \textbf{\bibinfo{volume}{13}},
  \bibinfo{pages}{4790--4794} (\bibinfo{year}{2013}).
\newblock \urlprefix\url{https://doi.org/10.1021/nl402407r}.
\newblock \bibinfo{note}{PMID: 24032432},
  \eprint{https://doi.org/10.1021/nl402407r}.

\bibitem{Bouhelier:2003}
\bibinfo{author}{Bouhelier, A.}, \bibinfo{author}{Beversluis, M.},
  \bibinfo{author}{Hartschuh, A.} \& \bibinfo{author}{Novotny, L.}
\newblock \bibinfo{title}{Near-field second-harmonic generation induced by
  local field enhancement}.
\newblock \emph{\bibinfo{journal}{Phys. Rev. Lett.}}
  \textbf{\bibinfo{volume}{90}}, \bibinfo{pages}{013903}
  (\bibinfo{year}{2003}).
\newblock
  \urlprefix\url{https://link.aps.org/doi/10.1103/PhysRevLett.90.013903}.

\end{thebibliography}
\end{document}

\title{Spatio-temporal sampling of near-petahertz vortex fields: supplemental document}

\author{Johannes Bl{\"o}chl}
\affiliation{Department of Physics, Ludwig-Maximilians-Universit\"at Munich, D-85748 Garching, Germany}
\affiliation{Max Planck Institute of Quantum Optics, D-85748 Garching, Germany}
\author{Johannes Sch{\"o}tz}
\affiliation{Department of Physics, Ludwig-Maximilians-Universit\"at Munich, D-85748 Garching, Germany}
\affiliation{Max Planck Institute of Quantum Optics, D-85748 Garching, Germany}
\author{Ancyline Maliakkal} %need to check current affiliation
\affiliation{Department of Physics, Ludwig-Maximilians-Universit\"at Munich, D-85748 Garching, Germany}
\affiliation{Max Planck Institute of Quantum Optics, D-85748 Garching, Germany}
\author{Natālija Šreibere} %need to check current affiliations
\affiliation{Department of Physics, Ludwig-Maximilians-Universit\"at Munich, D-85748 Garching, Germany}
\affiliation{Max Planck Institute of Quantum Optics, D-85748 Garching, Germany}
\author{Zilong Wang}
\affiliation{Department of Physics, Ludwig-Maximilians-Universit\"at Munich, D-85748 Garching, Germany}
\affiliation{Max Planck Institute of Quantum Optics, D-85748 Garching, Germany}
\author{Philipp Rosenberger}
\affiliation{Department of Physics, Ludwig-Maximilians-Universit\"at Munich, D-85748 Garching, Germany}
\affiliation{Max Planck Institute of Quantum Optics, D-85748 Garching, Germany}
\author{Peter Hommelhoff}
\affiliation{Laser Physics, Department of Physics, Friedrich-Alexander-Universit\"at Erlangen-N\"urnberg, D-91058 Erlangen, Germany}
\author{Andre Staudte}
\affiliation{Joint Attosecond Science Laboratory, National Research Council of Canada and University of Ottawa, Ottawa, Ontario K1A0R6, Canada}
\author{Paul B. Corkum}
\affiliation{Joint Attosecond Science Laboratory, National Research Council of Canada and University of Ottawa, Ottawa, Ontario K1A0R6, Canada}
\author{Boris Bergues}
\affiliation{Department of Physics, Ludwig-Maximilians-Universit\"at Munich, D-85748 Garching, Germany}
\affiliation{Max Planck Institute of Quantum Optics, D-85748 Garching, Germany}
\author{Matthias F. Kling}
\email{kling@stanford.edu}
\affiliation{Department of Physics, Ludwig-Maximilians-Universit\"at Munich, D-85748 Garching, Germany}
\affiliation{Max Planck Institute of Quantum Optics, D-85748 Garching, Germany}
\affiliation{SLAC National Accelerator Laboratory, Menlo Park, CA 94025, USA}
\affiliation{Applied Physics Department, Stanford University, Stanford, CA 94305, USA}

\begin{abstract}
This document provides supporting information on the main article. We present not only additional data, but also investigate different aspects of our work theoretically. 
Among them, the response function as well as the intensity scaling. 
The additional data includes the data for spatio-temporal scans,  an intensity calibration, a dispersion scan, a carrier-envelope-phase scan as well as for a non-linearity scan.
\end{abstract}

\maketitle

\section{Spatio-temporal scan of orbital angular momentum beams}
\begin{figure*}[htb!]
 \centering
 \includegraphics[width=135mm]{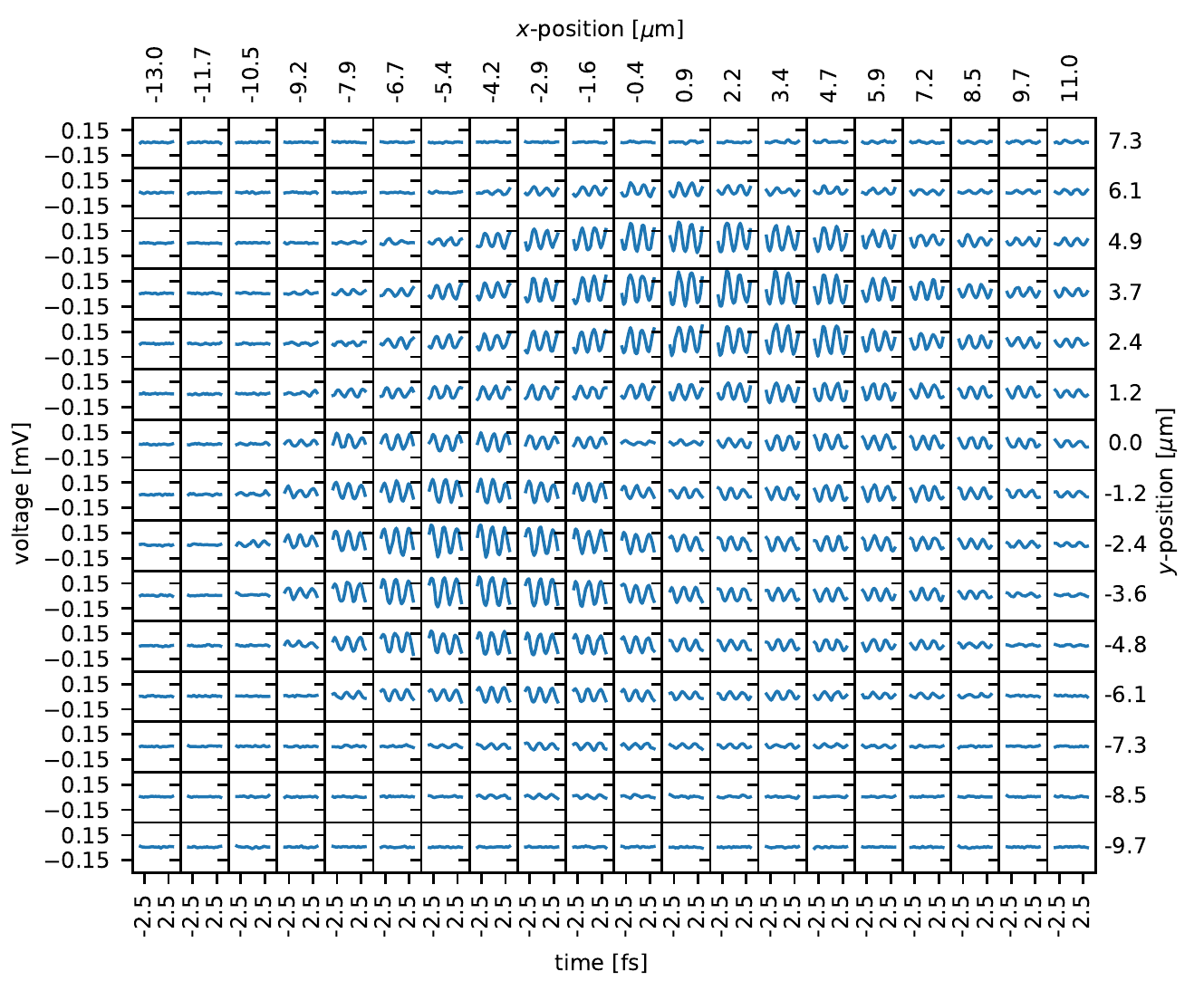}
 \caption[Spatio-temporal OAM scan]{Spatio-temporal scan of orbital angular momentum beams - At each point in the $x,y$-plane, we scanned over 2.5 optical cycles of the OAM beam. This measurement not only yields the phase information, but also the amplitude information, as we scanned over the maximum of the laserpulse. This is the raw data used to generate parts a), b), d) and g) of Fig.\,3 in the main text.}
 \label{fig:SI_spatTemp}
\end{figure*}

\begin{figure}[htbp!]
\centering
 \includegraphics[width=90mm]{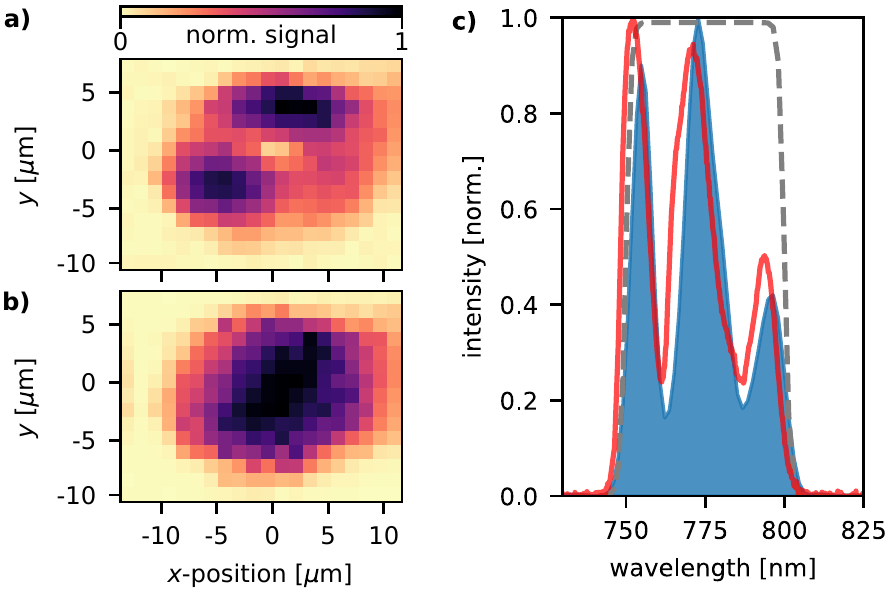}
 \caption[OAM Beam]{Amplitude and spectrum of the OAM beam: \textbf{a)} The amplitude of the current modulation in space has a ring shape. \textbf{b)} The corresponding injection current does not carry orbital angular momentum and is a round spot rather than a ring. \textbf{c)}  The blue filled area is the calculated spectrum of the nano\textsc{Tiptoe} measurement. It is located exactly in the region that the bandpass filter supports (dashed grey line). The spectrometer data (red) reproduces the peaks in the spectrum very well, but is slightly shifted in wavelength. This can be explained by the wavelength uncertainty of the device that is specified to be  2\,nm.}
 \label{fig:SI_OAMSpec}
\end{figure}

Nano\textsc{Tiptoe} is a reliable tool in sampling optical waveforms in space. 
Supplementary Figure\,\ref{fig:SI_spatTemp} shows the raw data used to generate Figs.\,3a,b,d and g) in the main text. 
For this purpose, the tip was scanned over the focal plane and the delay was sampled over 2.5 optical cylces of the orbital angular momentum beam at each point. 
Please note  that the detected signal was a voltage signal despite the fact that a current has been generated, as the transimpedance amplifier was converting the current to a voltage.
To each measured point, we fit a sine wave of the form \(u(t)=u_0\cdot\sin(2\pi f\cdot t + \varphi) +u_c\) to the data. 
There, $u_0$ is the amplitude, $f$ the frequency, $\varphi$ the phase and $u_c$ a constant offset that models the nonlinear distortions (see also SI Sec.\,\ref{sec:nonLinScan}).
The resulting image for \(\varphi(x,y)\) is shown in Fig.\,3d) in the main text. 
The offset \(u_c\) has been subtracted in main text Fig.\,3a). 
%The noisy points for the smallest $x$ and $y$ values can be explained with the low signal amplitude that is already close to the noise level.
Besides the phase information, we also got the amplitude information as we always scanned over the main pulse, as shown in main text Fig.\,3b) and SI Fig.\,\ref{fig:SI_OAMSpec}a). 
The current generated by the pump pulse is shown in part\,b) of SI Fig.\,\ref{fig:SI_OAMSpec}.
Despite the different mode sizes by a factor of 2.4, the comparison of SI Fig.\,\ref{fig:SI_OAMSpec}a) and b) suggests that probe and pump beam have an equal size.
However, this is misleading as firstly, when changing a Gaussian beam of fixed waist to an OAM beam, the apparent focal size becomes larger,  as the Gaussian function is multiplied by a factor \(\sim r^{|1|}\) for \(l=1, p=0\), c.f. \eqref{eq:GaussianBeam}.
And secondly, the detected signal is proportional to the field (part a)), whereas the injection current (part b)) is proportional to the ionization rate, that is, a non-linear function of the field which makes the spot size occurring smaller.

As mentioned in the main text, we can of course measure the complete electric field of the laser pulse. 
To prove that we indeed resolve the complete spectrum, we compared the calculated spectrum of the nano\textsc{Tiptoe} measurement to the spectrum taken with a commercial spectrometer (Thorlabs CCS200/M) and an integrating sphere, as can be seen in SI Fig.\,\ref{fig:SI_OAMSpec}. 
The spectrum of the nano\textsc{Tiptoe} data is exactly as broad as the bandpass filter supports. 
The spectrometer data shows the same peaks in the spectrum with similar height, but is slightly blue-shifted. 
This can be explained by the wavelength precision that the device has, as there are up to 2\,nm tolerance according to the producer. 
This is around the observed difference of 2-3\,nm.
For the OAM data, we averaged the two full-time scans performed at points A and B (c.f. Fig.\,2 in the main part).
The average was performed, as the integrating sphere is also averaging over the mode. 
However, we found that the difference in the spectra between point A and B is rather small.
A complete scan over the mode, especially also without bandpass filter would be very interesting regarding the wavelength dependent properties of the OAM generation.
However, such a measurement would be practically hard to implement under long-term controlled conditions, as with the scan times in our experiments the duration would be of the order of a day.

%\section*{Mode size estimation}
%A harder focusing with a 2`` OAP onto the nanometric needle tip provided smaller focal scales to probe.
%However, the OAM mode suffered from astigmatism that could not be prevented, see Fig.\,\ref{fig:SI_Linescan}a).
%Nevertheless, we compare the measured mode shape to a theoretical prediction while assuming that the abberations (or point-spreak function of our imaging system) will only increase the mode size, but never make the mode size smaller. 
%
%
%The focal spot size of the laser beam was measured using an imaging system with a plano-convex lens and a long focal length (0.4\,m). 
%This image provides a deviation factor of the spot size obtained from a calculation of perfect Gaussian optics using only the mode diameter and the focal length. 
%As a comparatively long focal length has been used to determine the deviation factor, the majority of deviations is expected to be defined by the laser mode shape and not by the imaging setup itself. 
%Therefore, this estimation using Gaussian optics plus an additional deviation factor sets a lower limit for the focal spot size in the experiments, where additional astigmatism played a role.
%For the dataset discussed next, we theoretically expect a beam waist \(w_0\) as defined below of \(w_0=2.7^{+0.7}_{-0.5}\,\mu\)m. 
%The uncertainty interval originates from an minimum/maximum value calculation.

For the expected field distributions of the Gaussian Beams and OAM beams, we used an expression for the electric field in space adapted from Ref.\,\cite{Allen:1992} which reads
\begin{equation}
\begin{split}
 E^l_p(r,\varphi,z,t)&=E(t)\frac{C(l,p)}{\sqrt{1+z^2/z_\text{R} ^2}}\left[ \frac{r\sqrt{2}}{w^2(z)}\right]^{|l|}
 L^l_p\left(\frac{2r^2}{w^2(z)}\right)\\
 &\times\exp\left[-\frac{r^2}{w^2(z)}+\frac{-ikr^2z}{2(z^2+z_\text{R}^2)}\right]\\
 &\times\exp\left[ - il\varphi + i(2p+l+1)\arctan\left(\frac{z}{z_\text{R}}\right)\right] \\
&\times \exp\left[-ikz +i\omega t\right].\label{eq:GaussianBeam}
\end{split}
\end{equation}
Here, \(r,\varphi\) and \(z\) are cylindrical coordinates, \(E(t)\) is the temporal envelope, \(z_\text{R}\) is the Rayleigh length, \(w(z)\) is the beam waist, \(L^l_p(.)\) are associated Laguerre polynomials and \(C(l,p)=\sqrt{\frac{2p!}{\pi (p+|l|)!}}\) is a constant defined by the azimuthal and radial indices \(l\) and \(p\), respectively. 
The central frequency is \(\omega\) and the \(k\) is the wavenumber.
In most cases, we assumed to be directly in the focal plane, i.e. we set \(z=0\) and \(w(z=0)=w_0\).
For a Gaussian beam, we set \(l=p=0\), whereas for our OAM beam, we set \(l=1, p=0\).
%Exemplary time-averaged field amplitudes can be seen in SI Fig.\,\ref{fig:SI_OAMSpec}a).  

Please note, we often found that the measured phasefront exhibits a linear behavior in space that can be quite large. It turned out to be that this was caused by imperfectly centered modes of pump and probe beam.
In an illustrating picture, this can be explained by the \(k\)-vectors of the pump and signal beam propagating under an angle through the focus which ends up in a linear phase relation in space between the two beams. This mode mismatch could be caused by parallel but slightly shifted beams.
In the measurements presented here, we prevented this additional phaseterm by carefully aligning not only parallelism of the beams, but also on centering one beam into the other.
The small additional curvature of the phase front we detected in the vortex beam measurements (c.f. main text Fig.\,3d)) was caused by a slightly quenched optic in the pump beam path and therefore even present without vortex plate.

\section{Dispersion scan}
\begin{figure*}[htbp!]
 \centering\includegraphics[width=135mm]{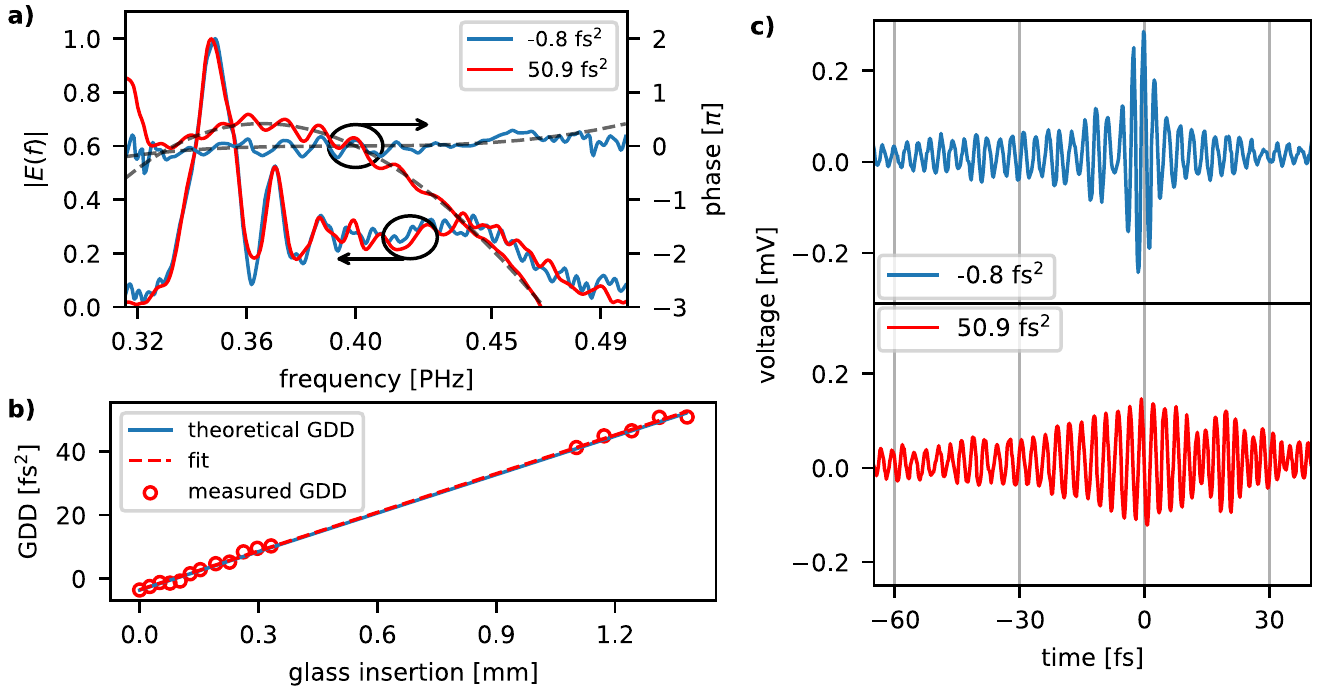}
 \caption[Dispersion scan with nano\textsc{Tiptoe}]{Dispersion scan with nano\textsc{Tiptoe}: \textbf{a)} The calculated spectra and phases for two measured pulses with different dispersion show the same spectral amplitudes but different quadratic phases. A fit of a third-order polynomial, weighted with the spectral amplitudes, delivers the value for the group-delay-dispersion.  
 The result of a systematic scan of the dispersion in the signal arm is shown in \textbf{b)}, clearly following a linear trend. A linear fit yields a GVD of 40.59\,fs$^2$/mm, which is only 0.63\% deviation from the literature value of 40.335\,fs$^2$/mm\,\cite{Malitson:1965} \textbf{c)} Time-domain signals corresponding to the spectra shown in a).}
 \label{fig:SI_DispScan}
\end{figure*}

To verify that nano\textsc{Tiptoe} probes the light field, we systematically scanned different light properties as the dispersion.
Therefore, we moved the fused silica wedges in the signal arm while keeping the pump beam untouched, i.e. compressed.
The temporal overlap was realigned using a manual stage that the piezo stage was attached to.
Please note, that neither bandpass filter nor vortex plate was used here.
Additionally, a fused-silica window (thickness \((1.05\pm0.05)\)\,mm) could be added to increase the scan range.
As can be seen in SI Fig.\,\ref{fig:SI_DispScan}a), the calculated spectrum from the data is similar for all glass insertions, whereas the time-domain pulse shapes are either chirped or compressed, see SI Fig.\,\ref{fig:SI_DispScan}c). 
The spectral phase, however, clearly shows a quadratic behavior when the pulse is chirped. 
Additionally, there are some modulations on the phase that are most likely caused by the 8 pairs of double angle chirped mirrors (PC70, Ultrafast Innovations) used for the pulse compression.
To extract the group-delay-dispersion (GDD), we fit a third-order polynomial to the spectral phase. 
We weighted the fit with the spectral amplitudes in order to avoid overfitting of phase terms with low amplitude. 
The resulting second order dispersion is shown in SI Fig.\,\ref{fig:SI_DispScan}b).
It clearly follows a linear scaling and a fit yields a dispersion coefficient of 40.59\,fs$^2$/mm, which is only 0.63\% deviation from the literature value of 40.335\,fs$^2$/mm\,\cite{Malitson:1965} for fused silica at 750\,nm central wavelength (0.4\,PHz).

\section{Response function of nanoTIPTOE}\label{sec:response}
\begin{figure*}[htbp!]
	\centering\includegraphics[width=135mm]{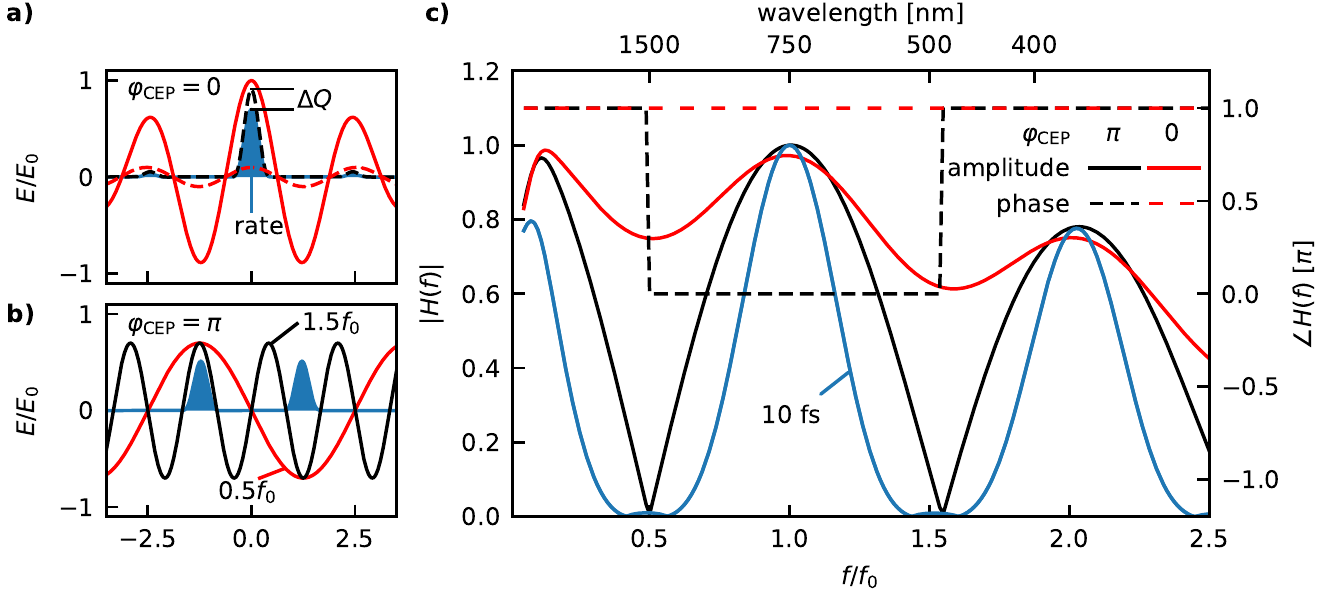}
	\caption[CEP-dependence of the nano\textsc{Tiptoe} response function]{\label{Nanotip_current_TipToeResponse} The nano\textsc{Tiptoe} response function: \textbf{a)} The nano\textsc{Tiptoe}-principle: The ionization rate (blue line area) of a strong pump pulse (red line) is perturbed by a weak signal pulse (red dashed line). The change in ionization is essentially proportional to the perturbing electric field at the time of the emission peak. At a surface, electron emission only occurs if the electric field points into the surface, that is, with \(\varphi_\text{CEP}=0\) in our coordinate system. \textbf{b)} Illustration of the reason for the minima in the amplitude response for \(\varphi_\text{CEP}=\pi\) at $0.5\cdot f_\mathrm{0}$ (red line) and $1.5\cdot f_\mathrm{0}$ (black line). \textbf{c)} Amplitude (solid lines) and phase (dashed lines) of the response function for \(\varphi_\text{CEP}=0\) (red) and \(\varphi_\text{CEP}=\pi\) (black) for a pulse width of 4.5\,fs (FWHM) calculated using the Fowler-Nordheim rate. For a 10\,fs pump pulse (blue line), practically no CEP-dependence is observed (intensity $3\cdot10^{13}\mathrm{W/cm}^2$, work function=4.5\,eV). }
\end{figure*}

The response function of nano\textsc{Tiptoe} consists of two major contributions: 
Firstly, the incident fields are enhanced due to the geometric shape of the tip and secondly, the ionization process at a surface which is the field sampling process. 
The response function of the latter has already been discussed in many details in Ref.\,\cite[][supplementary]{Bionta:2021}, where the \textsc{Tiptoe} technique has been applied to nano triangle arrays. 
We will therefore give a brief overview only.
%According to that reference, nano\textsc{Tiptoe} also allows to sample the same bandwidth as the pump pulse supports and additionally the first integer harmonics of the fundamental frequency. 
%However, when the signal pulse is at an half-integer harmonic of the pump pulse, the response amplitude is zero. 
%There, the situation is met, where the modulation of one ionization burst is canceled by the modulation of another one. 
%One expectation are single cycle driving pulses: If only a single ionization burst exists, the current modulation always follows the perturbing field, as long as its frequency is not higher than twice the inverse width in time of the burst.
%In total, no modulation is observable.
\textsc{Tiptoe}-like methods rely on a few-cycle pump pulse that ideally only ionizes some medium during the strongest half-cycle creating a field-dependent current.
The total generated charge, i.e. the time-integral over the ultrafast current is then detected.  
This charge is now modulated by a weak perturbing field, the signal field. 
As this modulation happens only during one sub-cycle ionization burst, the modulation will follow the signal field, depending on the temporal delay between pump and signal field\,\cite{Park:2018}, see SI Fig.\,\ref{Nanotip_current_TipToeResponse}a). 
We can write the emitted charge \(Q\) in terms of the ionization rate \(w(t)\) and the delay \(\tau\)\,\cite{Park:2018,Bionta:2021}:
\begin{equation}
	\label{Eq_exp_TIPTOEfull}
	Q(\tau)\propto \int  w\Big(E_\text{p}(t-\tau)+E_\text{s}(t)\Big) \mathrm{d}t,
\end{equation}
where the time integral extends over the pulse length. Since the signal field is weak, the kernel of the integral can be Taylor-expanded up to first order:
\begin{equation}
	\label{Eq_exp_TIPTOE}
	Q(\tau) \propto \int w\Big( E_\mathrm{p}(t-\tau)\Big)+\frac{\mathrm{d}w}{\mathrm{d}E}\Bigg|_{E_\mathrm{p}(t-\tau)}\cdot E_\mathrm{s}(t) \mathrm{d}t,
\end{equation}
where the last term corresponds to the cross-correlation of the gating function with the signal field\,\cite{Bionta:2021}. The first term corresponds to the rate without signal field. Moreover, the gating function for an tunneling emission burst at $t_0$ can be approximated as a delta-function $\delta(t-\tau-t_0)$, yielding an expression for $\Delta Q$:
\begin{equation}
	\Delta Q(\tau)\propto \int \frac{\mathrm{d}w}{\mathrm{d}E}\Bigg|_{E_\mathrm{p}(t-\tau)}\cdot E_\mathrm{s}(t)\, \mathrm{d}t \propto \frac{\mathrm{d}w}{\mathrm{d}E}\Bigg|_{E_\mathrm{p}(t_0)}\cdot E_\mathrm{s}(t_0+\tau).
\end{equation}
For a single emission burst, the charge modulation is thus approximately proportional to the electric field. For several tunneling emission bursts, the last expression on the left hand side becomes a sum over all burst times $t_{k}$.
This is illustrated in Fig.\,\ref{Nanotip_current_TipToeResponse}\,b) for a pump pulse (not shown) with \(\varphi_\text{CEP}=\pi\), where two dominating equal emission bursts occur. 
As can be seen, in this case, signal frequencies at $0.5\,f_0$ (red line) and $1.5\,f_0$ (black line), where $f_0$ is the pump pulse frequency, do not lead to a modulation of the emitted charge. 
The reason is that the emission modulation of the first burst is counteracted by the second burst since the signal fields point into opposite direction. This is not the case for \(\varphi_\text{CEP}=0\).
Please note, that in our coordinate system, the tip is pointing in negative \(x\)-direction.
Therefore, a pulse with \(\varphi_\text{CEP}=0\) will cause a single ionization burst, as its strongest half-cycle points into the tip surface.
The resulting CEP-dependence of the spectral response function can be seen in Fig.\,\ref{Nanotip_current_TipToeResponse}c), and has been extensively discussed in Ref.\,\cite{Bionta:2021}. 
The spectral response function can be calculated either by Fourier transforming the gating function in Eq.\,\ref{Eq_exp_TIPTOE}, or by evaluating the full expression for the emitted charge for a given pump and signal field in Eq.\,\ref{Eq_exp_TIPTOEfull} subsequently forming the ratio of $\Delta Q$ and $E_\mathrm{s}$ in the Fourier-domain. 
We chose the latter approach, since it does not rely on the first-order Taylor-expansion. 
The results shown here use a Fowler-Nordheim emission rate\cite{Fowler:1928,Piglosiewicz:2014} assuming a work function of 4.5\,eV, a pump pulse at an intensity-FWHM of 4.2\,fs, a central wavelength of 750\,nm and intensity of approximately $3\cdot10^{13}\,\mathrm{W/cm}^2$.  
In order to obtain a large bandwidth, we chose a signal pulse duration of 2\,fs at 750\,nm wavelength and a relative field strength of $10^{-5}$.

Figure\,\ref{Nanotip_current_TipToeResponse}c) shows the amplitude (solid lines) and phase (dashed lines) of the spectral response function \(H(f)\) for a CEP of 0 (red) and $\pi$ (black), respectively. 
As discussed above, for \(\varphi_\text{CEP}=\pi\), the amplitude response exhibits zeros at $f=0.5\,f_0+ n\cdot f_0$, where $n$ is an integer. 
The response function for \(\varphi_\text{CEP}=0\) only shows a slight modulation caused by the small satellite emission bursts one cycle earlier and later as shown in Fig.\,\ref{Nanotip_current_TipToeResponse}a). 
Around $f_0$, the phase of the response function changes with the CEP of the pump pulse. 
As pointed out in Refs.\,\cite{Bionta:2021,Liu:2021,Liu:2022}, if both pump and signal pulse originate from the same laser source, this allows the auto-characterization of the pulse without CEP-stabilization, since the relative phase of both pulses does not change. 
However, this statement only holds true if the bandwidth is relatively narrow. 
%We have investigated that the change of the amplitude of the response function $|H|$ with the CEP can be measured with broadband pulses (see below). 
%For comparison, we also show the $|H|$ for a pump pulse with the same parameters as above, but a pulse duration of 10\,fs (solid blue line). 
%In the latter case, the CEP-dependence is not visible anymore.
For very broadband laser pulses, the CEP dependence of the response amplitude is expected to play a role. 
However, in our experiments we found only minor influence of the CEP of the driving pulse on the detected spectrum.

\begin{figure}[htbp!]
 \centering
 \includegraphics[width=89mm]{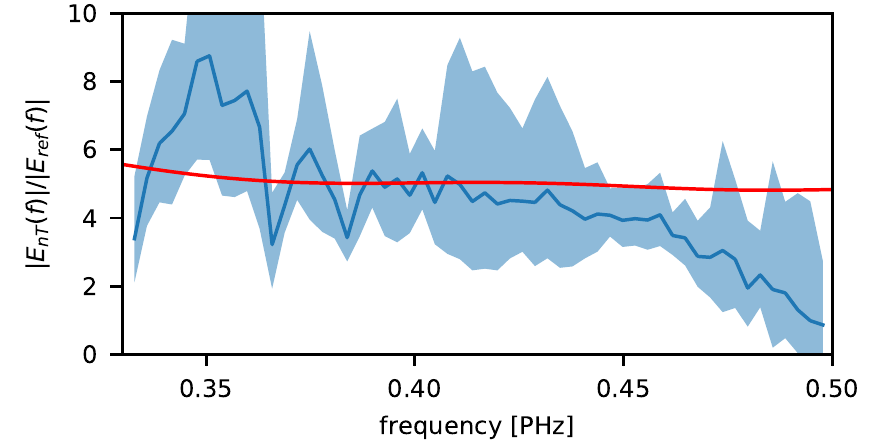}
 \caption[Response]{Response function - The extracted response \(|H|=\frac{|E_\text{nano\textsc{Tiptoe}}|}{|E_\text{ref}|}\) compared to FDTD simulations (red line) with a 4.5\,fs Gaussian beam.}
 \label{fig:SI_ntSpec}
\end{figure}

Additional to these properties inherited by the ionization process, the near-field enhancement plays in.
The reference measurement presented in main text Fig.\,2 allows to calculate the spectral response function as shown in SI Fig.\,\ref{fig:SI_ntSpec}.
The function shown there is the difference in response between nano\textsc{Tiptoe} and the reference, as the reference might still have a (potentially negligibly flat) response. 
Note, that we can only calculate the response function up to a scaling factor, as the amplitudes of the data are not comparable and were normalized first (main text Fig.\,2).
For comparison, we simulated the response function of the enhancement (red line in SI Fig.\,\ref{fig:SI_ntSpec}) using a commercial finite-difference time-domain (FDTD) solver (Lumerical FDTD).
We simulated a nanotip modeled by a truncated cone with \(10.5^\circ\) half-opening angle and a half-sphere with radius \(15\)\,nm as termination and a Gaussian pulse with 4.5\,fs duration.
The simulated response function is rather flat over the spectral region of interest, but agrees with the nano\textsc{Tiptoe} data within the uncertainty range.

\section{CEP dependence of nanoTIPTOE}
Recent literature suggests that \textsc{Tiptoe}-type measurements only detect the relative phase between the pump and signal pulse\cite{Bionta:2021,Liu:2021,Liu:2022}. 
This corresponds to a phase term of the response function that is proportional to the carrier-envelope-phase of the pump pulse\,\cite{Bionta:2021}.
For a more detailed discussion, see also section\,\ref{sec:response}.
We experimentally confirmed that indeed the relative phase is detected. 
However, phase changes in one arm of the interferometer were indeed measurable which not only establish the applicability of nano\textsc{Tiptoe} even to CEP-unstable lasers, but allow to detect modifications in one interferometer arm like the vortex generation.

\begin{figure*}[htbp!]
\centering
\includegraphics[width=135mm]{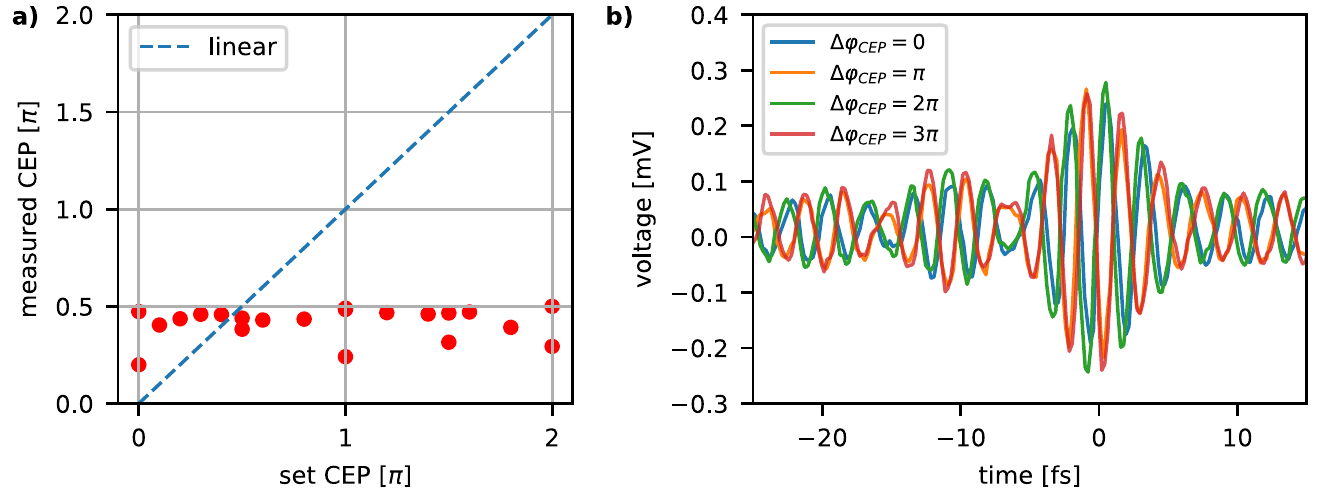} % need to redo this figure to fit into two column layout
\caption[CEP Dependence]{CEP dependence: \textbf{a)} The measured CEP of the nano\textsc{Tiptoe} signal is nearly constant (red dots), whereas the set CEP increases from 0 to \(2\pi\). \textbf{b)} When only changing the CEP of the signal beam, the phase change can be measured well. The plot is showing raw data.}
\label{fig:SI_CEP}
\end{figure*}

For the investigations on the CEP dependence of both pulses, we changed their CEP simultaneously using the Dazzler in the laser amplifier. 
SI Figure\,\ref{fig:SI_CEP}a) illustrates that the measured CEP with nano\textsc{Tiptoe} is nearly constant even though the carrier-envelope-phases of pump and signal pulse changes. 
If the response phase was independent of the CEP of the pump pulse, a linear increase would be expected (dashed line in SI Fig.\,\ref{fig:SI_CEP}a)).

To verify the phase sensitivity, we changed the CEP of the signal pulse only using fused silica wedges.
Knowing the group and phase refractive index of \(n_\text{ph}=1.4542\) and \(n_\text{gr}=1.4689\)\,\cite{Malitson:1965}, we calculated the needed thickness change of the glass for a \(\pi\) phase flip to be \(x=\frac{\pi}{(2\pi/\lambda)\cdot(n_\text{gr}-n_\text{ph})}=25\,\mu\)m. 
This change corresponds to a displacement of \(730\,\mu\)m at \(2^\circ\) opening angle of the wedge.
From the variable \(x\), we calculated the group-delay and shifted the signal by the corresponding time-shift of \(38.5\)\,fs.
We employed the full scanning range of the piezo delay stage which allows us to directly shift the signal by the time shift. 
Fig.\,\ref{fig:SI_CEP}b) shows that the unfiltered data after consecutive CEP shifts indeed shows a phase flip of \(\pi\) each time.

\section{Intensity calibration}
For the incident intensity, we found a power to intensity conversion given by
\begin{equation}
\begin{split}
 I(P[\text W])&=4.2\times 10^{15}\cdot P[\text{W}]\frac{\text W}{\text{cm}^2} \\
& \pm \left[ 0.9\times 10^{15}\cdot P[\text{W}]+4.2\times10^{15}\cdot \Delta P[\text{W}] \right]\frac{\text W}{\text{cm}^2}\label{eq:incidentIntens4inch}
\end{split}
\end{equation}
for the pump beam with 4'' focal length.
%, and 
%\begin{align}
 %I(P[\text W])=1.7\times 10^{20}\cdot P[\text{W}]\frac{\text W}{\text{m}^2} \pm \left[ 0.3\times 10^{20}\cdot P[\text{W}]+1.7\times10^{20}\cdot \Delta P[\text{W}] \right]\frac{\text W}{\text{m}^2}
%\end{align}
%for 2`` focal length. 
The incident power \(P\) has the measurement tolerance \(\Delta P\).
The above formula follows from an estimation via Gaussian optics.
However, we corrected deviations by an image of the focal spot on a camera in a separate imaging setup.
Accordingly, the values given above are only \((30\pm5)\%\) of the expected intensity by pure Gaussian optics.
The given errors follow from a Gaussian error propagation of the uncertainties of all the initial quantities. 
The signal beam intensity is larger by a factor of around 5.9 as the mode size is different.
Taking also the lower repetition rate into account, the intensity should be around 11.8 times higher for the same incident power.

\begin{figure}[htbp!]
 \centering
 \includegraphics[width=89mm]{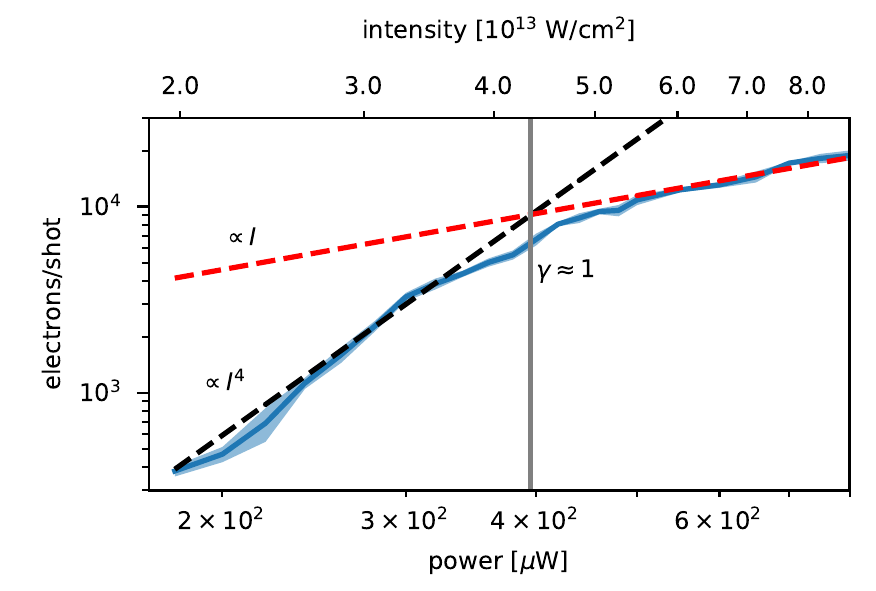}
 \caption[Intensity Scan]{Intensity scan: The number of emitted electrons (solid blue line and blue area) scales as \(I^4\) (dashed black line) for low (local) intensities, and as \(I^1\) (dashed red line) for higher (local) intensities. The dashed lines are only a guide for the eye. }
 \label{fig:SI_intensityScan}
\end{figure}

To estimate the local intensity at the tip, we investigated the intensity dependence of the emission current. 
For this purpose, the pump beam power is scanned while detecting the current, see SI Fig.\,\ref{fig:SI_intensityScan}.
We estimated the number of emitted electrons per shot taking 10\,kHz repetition rate, an amplification of \(10^9\,\frac{\text V}{\text A}\) and the damping rate, as discussed in the main text, into account.
The ionization current exhibits two different scaling laws:
For low incident intensity, the current scales as \(I^4\), whereas for higher intensities, the scaling is proportional to \(I\).
Similar curves have been observed in previous work\,\cite{Bormann:2010,Yalunin:2011,Piglosiewicz:2014,Swanwick:2014,Krueger:2018:2,Garg:2020}.
We follow a similar aproach as in those references, especially Ref.\,\cite{Bormann:2010} suggests the identification of the kink in the nonlinearity with the transition from multiphoton to tunneling ionization. 
This allows to set the Keldysh parameter \(\gamma=\frac{\omega\sqrt{2m_\text e \Phi}}{eE}\) to unity at that point\,\cite{Keldysh:1965}, which sets the enhanced intensity knowing the work function of tungsten of \(\Phi=4.5\)\,eV\,\cite{Michaelson:1977}.
Even within different measurements with the same nanometric needle tip in different beam times, the kink in nonlinearity was always at an incident power of \(P=(3.9\pm0.4)\times10^2\,\mu\)W for the 4'' focal length optics.
% and at \(P=(1.5\pm0.2)\,\mu\)W for the 2`` focal length OAP.
Regarding the nonlinearity in the multiphoton regime, the scaling of \(\propto I^4\) is of higher order than expected from the number of photons needed to ionize the tip, that is 3 at a photon energy of 1.65\,eV, see Ref.\,\cite{Mainfray:1991}. 
Similar observations have been made for instance in Ref.\,\cite{Bormann:2010}, as well as in other work on tungsten\,\cite{Fujimoto:1984,Nicolaou:1975,Modinos:1976,Christensen:1974,Feuerbacher:1974,Lee:2004}.
In those references, a higher density of states further below the Fermi level causes the higher nonlinearity.
Having fixed the kink postion, we can write the enhanced intensity as
\begin{equation}
\begin{split}
 I(P[\text W])&=1.1\times 10^{17}\cdot P[\text{W}]\frac{\text W}{\text{cm}^2} \\
&\pm \left[ 1.2\times 10^{16}\cdot P[\text{W}]+1.1\times10^{17}\cdot \Delta P[\text{W}] \right]\frac{\text W}{\text{cm}^2}
\end{split}
\end{equation}
for the pump beam.
% with 4'' focal length, and 
%\begin{align}
 %I(P[\text W])=2.9\times 10^{21}\cdot P[\text{W}]\frac{\text W}{\text{m}^2} \pm \left[ 3.8\times 10^{20}\cdot P[\text{W}]+2.9\times10^{21}\cdot \Delta P[\text{W}] \right]\frac{\text W}{\text{m}^2}
%\end{align} 
%for 2`` focal length. 
For 1\,mW incident power, the enhanced intensity at the tip is \(I=(1.1\pm0.2)\times10^{14}\,\frac{\text W}{\text{cm}^2}\). 
Comparing this to the incident intensity calibration above, we can estimate the intensity enhancement to \(26^{+14}_{-9}\), or equivalently, a field enhancement of \(5.1^{+1.2}_{-0.9}\).
We note that Ref.\,\cite{Schoetz:2021:spaceChargeNanophot} has demonstrated that the kink in the photoemission scaling is also connected to the onset of charge interaction. However, it is also stated there that, despite there being charge interaction, the measurement of field-dependent currents is still possible. Accordingly, we conclude that space-charge interaction might be present in our experiments, but does not affect the sampled waveform. 

The estimated enhanced near-field intensities in the nano\textsc{Tiptoe} regime were \((7.2\pm0.9)\times10^{13}\)\,W/cm$^2$ and \((4.6\pm0.6)\times 10^{12}\)\,W/cm$^2$ for pump and signal beam, respectively, for the dataset presented in Fig.\,2 in the main text. 
The incident intensities were lower according to the enhancement factor.
In the  \textsc{Tiptoe} reference measurements, the incident intensities were around \((4\pm1)\times10^{13}\)\,W/cm$^2$ for the pump beam and \((2.0\pm0.4)\times10^{11}\)\,W/cm$^2$ for the signal beam.
The \textsc{Tiptoe} mechanism relies on the signal field being only a weak perturbation. Despite the signal beam intensity in our experiments being more than 6\% of the pump intensity, we did not find significant non-linear distortion, see also the experimental investigation in section\,\ref{sec:nonLinScan}.

\section{Different regimes of nanoTIPTOE}\label{sec:nonLinScan}
\begin{figure*}[htbp!]
	\centering
	\includegraphics[width=135mm]{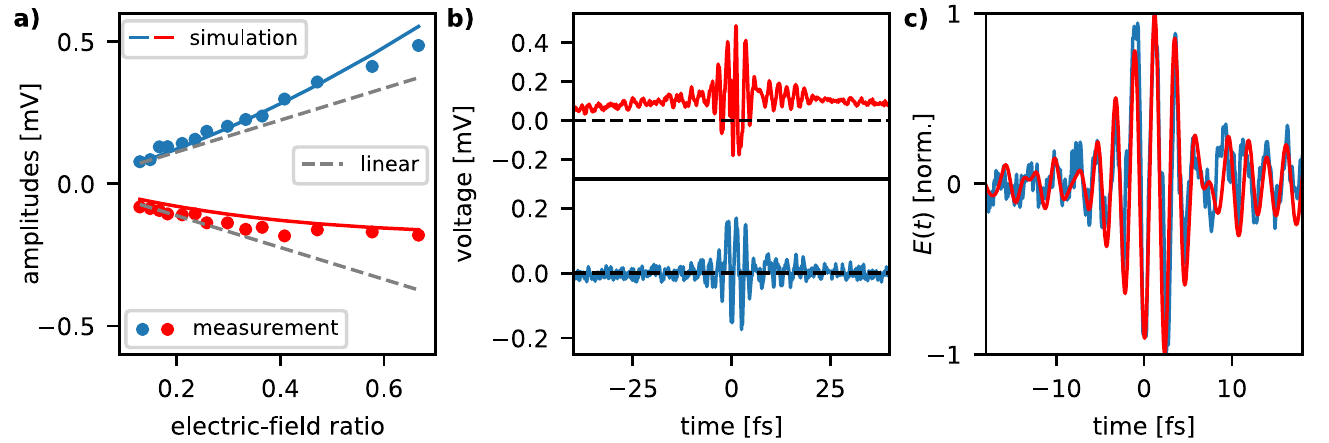}
	\caption[Nonlinearity Scan]{\label{fig:SI_powerscan} Nonlinearity scan: \textbf{a)} The amplitudes in positive and negative direction of the sampled waveform (red and blue dots) leave the linear scaling regime (dashed) at a field-ratio of around 0.3. The behavior follows the same nonlinearity as expected by simulations modeling the process with Fowler-Nordheim tunneling. \textbf{b)} Two waveforms with taken with a different signal field strength show a low-frequency background as well well as harmonic distortions if the field strength is higher (red line). \textbf{c)} The blue line represents the raw data sampled with low signal field strength. It is well reproduced when applying a Fourier filter between 0.3 and 0.6\,PHz to the waveform sampled with higher field strength (red).  } 
\end{figure*}

Originally, \textsc{Tiptoe} has been derived for conditions where the signal field only constitutes a small perturbation of the pump field\,\cite{Park:2018}, but has also been extended beyond this condition at the expense of a more complicated retrieval\,\cite{Cho:2019,Cho:2021}.
We experimentally verified that the linear approximation (see eq.\,\eqref{Eq_exp_TIPTOE}) in the nano\textsc{Tiptoe} sampling process breaks down if the field ratio between signal and pump beam is too large, see SI Fig.\,\ref{fig:SI_powerscan}a). 
The scaling of the detected amplitudes leaves the linear regime as soon as the electric field ratio is larger than 0.3.
To deduce the actual field-ratio, we employed similar mode sizes of signal and pump beam in order to avoid the uncertainties that come with the intensity calibration above.
By doing so, we directly extract the field-strength ratio by the square-root of the power ratio, where we varied the signal field while keeping pump constant. 
Here, we made sure that the dispersion of both beams is balanced best.
The results reproduce simulations using a Fowler-Nordheim tunneling\,\cite{Fowler:1928} rate with comparable field-strengths as in our experiments, but with every second optical half-cycle suppressed\,\cite{Yalunin:2011}.

At the highest signal powers,  we observe a prominent asymmetry of the nano\textsc{Tiptoe} signal that is characterized by a low-frequency offset, reminiscent of an autocorrelation signal, see SI Fig.\,\ref{fig:SI_powerscan}b) upper graph.
In contrast, for lower signal field strength, no such offset is visible. 
We refer to that offset as non-linear offset or non-linear distortion, as it is only present when nano\textsc{Tiptoe} has left the linear regime.
This low frequency offset can be seen in terms of higher orders of the Taylor expansion in \eqref{Eq_exp_TIPTOE}. 
There, the second order term is proportional to \(E_\text{S}^2(t)\) which is the time-domain representation of sum and differency frequency generation\,\cite{Park:2018}.

We also demonstrate that using Fourier transform filtering of a dataset with too large perturbation, we can reconstruct the sampled waveform taken with appropriate field strength, see SI Fig.\,\ref{fig:SI_powerscan}c).
There, the waveform shown in SI Fig.\,\ref{fig:SI_powerscan}b) (red), has been Fourier transformed and filtered.
Frequency components below 0.3 and above 0.6\,PHz have been cut. 
The normalized back-transformation agrees well with another waveform sampled with a smaller perturbation of the ionization. 
As we employed a few-cycle pump pulse, a more sophisticated reconstruction algorithm\,\cite{Cho:2021} is not necessary. 
%For a deeper understanding of the photoemission process and the signal formation under our conditions, the dependence of the photocurrent on signal and pump intensity is investigated. The current signal, demodulated at the repetition rate, which is proportional to the total photoemission current versus the pump power is shown in Fig.\,\ref{Nanotip_current_intensity_dependence}\,a). Here, the number of detected electrons per shot (blue line) has been calculated from the nominal transimpedance and the total number of electrons should be slightly higher, as mentioned above. The thickness of the line indicates the uncertainty. The signal arm has been blocked. When increasing the pump power from 70\,$\mu$W to 300\,$\mu$W (incident intensity from 0.8 to $3.3\cdot10^{12}\mathrm{W/cm}^2$) an increase of the number of detected electrons by almost two orders of magnitude is observed from below 10 up to around 1000 per shot. At around 140\,$\mu$W (14 nJ/pulse), there is a noticeable change of the slope of the curve, which is equivalent to the nonlinearity of the photoemission process. The nonlinearity (gray triangles), calculated using finite differences between neighboring datapoints, changes from around 4.2 to roughly 1.8, as confirmed by fitting the slope of the measured photoemission current over the region indicated by the black lines.

\section{Damage threshold for nanoTIPTOE} 
Similar intensities have been reached in other needle tip experiments\cite{Piglosiewicz:2014, Hoff:2017,Bormann:2010} without observing damage to the tip. Specifically Ref.\cite{Bormann:2010} states that under conditions similar to our experiment, even at intensities up to a factor 6 above the kink, no damage seems to occur to the tip. Several studies indicate that one of the major damaging mechanisms for nanostructures is heating by the electric field inside the material in combination with inefficient heat conduction\cite{Morgner2013NJPXUVnanoantennaDamage, Liu2013DamageAgNanowires, Summers2014NWdamage}. Accumulated heating is rather a problem at MHz-repetition rates\cite{Morgner2013NJPXUVnanoantennaDamage} which has also been shown to potentially alter the emission characteristics\cite{Bionta:2016}. In contrast, at kHz rates there is enough time between consecutive pulses for the heat to be dissipated. Moreover, since the nanometric needle tip quickly takes on a micron length scales away from the apex due to the conical structure of the shank, a better heat dissipation can be expected compared to e.g. nanowires. Nevertheless, those nanowires exhibit a damage threshold of around $10^{13}\mathrm{W/cm^2}$ of incident intensity at kHz repetition rates\cite{Summers2014NWdamage}. Finally, as a general rule the damage threshold increases with decreasing pulse length. In this regard, our experimental conditions where a tungsten tip is irradiated by sub-two-cycle laser pulses at 10\,kHz repetition rate and below, we do not expect damage as our incident intensities are even below \(10^{13}\,\mathrm{W/cm^2}\), as can be calculated from \eqref{eq:incidentIntens4inch}.
Indeed, we do not observe significant changes of the photocurrent over a several days long period of measurements.
We also conclude that there is no significant damage by the repeatability of the measurement with the same nanometric needle tip.

% Bibliography
%